\documentclass{llncs}
\usepackage{graphicx}
\usepackage{subfigure}
\usepackage[ruled,vlined,linesnumbered]{algorithm2e}
\usepackage{authblk}
\usepackage{amsmath}
\usepackage{amsfonts}

\usepackage[font=bf,skip=5pt]{caption}
\let\llncssubparagraph\subparagraph
\let\subparagraph\paragraph
\usepackage{titlesec}
\let\subparagraph\llncssubparagraph
\titlespacing\section{0pt}{12pt plus 4pt minus 2pt}{4pt plus 2pt minus 2pt}
\titlespacing\subsection{0pt}{10pt plus 4pt minus 2pt}{4pt plus 2pt minus 2pt}
\titlespacing\subsubsection{0pt}{8pt plus 4pt minus 2pt}{10pt plus 2pt minus 2pt}
\begin{document}


\title{Speed Partitioning for Indexing Moving Objects}

\author{Xiaofeng Xu\inst{1} \and Li Xiong\inst{1} Vaidy Sunderam\inst{1} \and Jinfei Liu\inst{1}\and Jun Luo\inst{2}}

\institute{Department of Math/CS, Emory University, Atlanta, GA, USA,\\
	\email{\{xiaofeng.xu,lxiong,vss,jinfei.liu\}@emory.edu},
	\and
	Shenzhen Institutes of Advanced Technology, Chinese Academy of Sciences\\
	\email{jun.luo@siat.ac.cn}}

\maketitle 

\begin{abstract}
Indexing moving objects has been extensively studied in the past decades. Moving objects, such as vehicles and mobile device users, usually exhibit some patterns on their velocities, which can be utilized for velocity-based partitioning to improve performance of the indexes. Existing velocity-based partitioning techniques rely on some kinds of heuristics rather than analytically calculate the optimal solution. In this paper, we propose a novel speed partitioning technique based on a formal analysis over speed values of the moving objects. We first show that speed partitioning will significantly reduce the search space expansion which has direct impacts on query performance of the indexes. Next we formulate the optimal speed partitioning problem based on search space expansion analysis and then compute the optimal solution using dynamic programming. We then build the partitioned indexing system where queries are duplicated and processed in each index partition. Extensive experiments demonstrate that our method dramatically improves the performance of indexes for moving objects and outperforms other state-of-the-art velocity-based partitioning approaches.
\end{abstract}

\section{Introduction}
\label{intr}
Over the past few decades, the rapid and continuous development of positioning techniques, such as GPS and cell tower triangulation, has enabled information to be captured about continuous moving objects, such as vehicles and mobile device users. \textit{Location-based services} (LBSs) and \textit{location-dependent queries} have become popular in modern human society \cite{DBLP:books/mk/schillerV04/SchillerV04}. Techniques for managing databases containing large numbers of moving objects and processing predictive queries \cite{DBLP:conf/sigmod/SaltenisJLL00} \cite{DBLP:conf/vldb/PapadiasTS03} have been extensively studied and are becoming increasingly important in order to support many emerging applications including real-time ride sharing (e.g. Uber) and location based crowd sourcing (e.g. Waze).

By storing timestamped locations, traditional database management systems (DBMSs) can directly represent moving objects \cite{DBLP:conf/stdbm/NascimentoST99}. However, this approach is impractical because most applications require high update rates in order to maintain the stored locations of the moving objects up to date. Therefore, motion functions are used instead, which significantly reduce the number of updates, for moving object databases (MODs) \cite{DBLP:conf/vldb/JensenP07} \cite{DBLP:conf/icde/SistlaWCD97}. Moreover, motion functions enable MODs to perform predictive spatio-temproal queries \cite{DBLP:conf/sigmod/SaltenisJLL00} \cite{DBLP:conf/vldb/PapadiasTS03} that retrieve near future locations of the moving objects. 

Indexes are used to improve query performance of MODs. Due to high update rate in real world applications, not only query performance but also update overhead must be considered while indexing MODs. Indexes for MODs in the literature can be categorized into tree-based indexes (e.g. \cite{DBLP:conf/sigmod/SaltenisJLL00} \cite{DBLP:conf/vldb/PapadiasTS03} \cite{DBLP:journals/vldb/SilvaXA09} \cite{DBLP:conf/vldb/JensenLO04} \cite{DBLP:journals/vldb/YiuTM08}) and grid-based indexes (e.g. \cite{DBLP:conf/sigmod/PatelCC04} \cite{DBLP:conf/gis/SidlauskasSCJS09} \cite{DBLP:conf/sigmod/SidlauskasSJ12} \cite{DBLP:journals/vldb/SidlauskasSJ14}). Typical tree-based indexes are balanced, i.e. the number of indexed objects within each leaf node is about the same. Therefore query performance of such structures can be estimated by the number of nodes accessed when processing a query \cite{DBLP:conf/vldb/PapadiasTS03}. The query performance of grid-based indexes depend on different factors, as the grid cells might contain quite different number of objects. In this work, we consider only tree-based indexes and leave grid-based ones for future work.                


In most real world applications, moving objects usually exhibit particular patterns on velocities (including speed values and directions). Therefore, velocity-based partitioning can be applied to the indexes to reduce performance deterioration caused by location proximity changes of the moving objects as time elapses. Zhang et.al. \cite{DBLP:journals/pvldb/ZhangCJOZ09} proposed the first idea of velocity-based partitioning for indexing moving objects. In their method, they first find $k$ velocity \textit{seeds} which maximize the \textit{velocity minimum bounding rectangle} (VMBR), then partition the moving objects by assigning them to the nearest seed. In this way, the moving objects are partitioned into $k$ parts and the VMBR for each part is minimized. Nguyen et.al. \cite{DBLP:journals/pvldb/NguyenHZW12} proposed another velocity-based partitioning technique that partitions the indexes based on directions of the moving objects. This method clusters the moving objects based on their distance to the so-called \textit{dominant velocity axes} (DVAs) in the velocity domain. This clustering strategy dramatically reduces the search space expansion when most of the moving objects move along DVAs. 

\subsection{Motivations}
In most real world scenarios, speed values of the moving objects are always characterized by both the nature of the moving objects and the environment. For example, pedestrian walking speeds for human beings range from 0 mph to 4 mph; driving speeds for vehicles in city road networks range from 0 mph to 100 mph; ground speeds for commercial airplanes usually range from 500 mph to 600 mph. Moreover, in most city road networks, speed values of the vehicles are also characterized by the categories of the roads. For example, most vehicles drive between 50-80 mph on highways, and 20-40 mph on street ways or even slower when the roads are busy.

This distribution of speed values of the moving objects can have significant impacts on query performance of the indexes. Query performance of typical tree-based indexes for MODs can be estimated by the average number of node accesses \cite{DBLP:conf/vldb/PapadiasTS03}. However, high speed moving objects will significantly enlarge the spatial areas of the index nodes containing them, which will likely incur unnecessary accesses to the low speed ones within the same nodes while processing queries. Thus partitioning the indexes by speed values of the moving objects can significantly improve query performance. Moreover, partitioning will reduce the number of objects in each index partition, which also helps accelerate update operations.  

\subsection{Contributions}
Motivated by above observations, we propose the novel speed partitioning technique. The proposed method first computes the optimal points (ranges) for partitioning, based on which the partitioned indexing system is built. On top of speed partitioning, an optional second-level partitioning, based on directions of the moving objects, is performed within each speed partition, which will further improve performance of the indexing system. Note that the location and speed distributions might change as time elapses that leads to changes on the optimal speed partitioning. Our proposed system can handle these changes through periodical partition update routines. Moreover, the speed partitioning technique is generic and can be applied with various tree-based indexes. 
Contributions of this paper can be summarized as follows:
\begin{itemize}
  \item We propose a novel method for estimating the search space expansion which can be used as a generic cost metric to estimate query performance of tree-based indexes for MODs. 
  \item We propose the novel speed partitioning technique which minimizes search space expansion of the indexes using dynamic programming.
  \item Extensive experiments show that our proposed approach prominently improves update and query performance of two state-of-the-art MOD indexes (the B$^x$-tree and the TPR$^\star$-tree) and outperforms other state-of-the-art velocity-based partitioning techniques.
\end{itemize}

The remainder of this paper is organized as follow. 
In Section \ref{sec:rela} we review the related works about tree-based indexes for MODs and velocity-based partitioning techniques. In Section \ref{sec:prob}, we introduce the concept of search space expansion and, based on which, we formulate the optimal speed partitioning problem. In Section \ref{sec:part}, we present the speed partitioning technique and the partitioned indexing system. Experimental studies are presented in Section \ref{sec:expe}. In Section \ref{sec:conc}, we conclude this paper and discuss some future work.

\section{Related Work}
\label{sec:rela}
In this section, we introduce some related work about tree-based indexes for MODs, which are extensions of the basic data structures of R-trees \cite{DBLP:conf/sigmod/Guttman84}, B$^+$-trees, and quad trees \cite{DBLP:journals/acta/FinkelB74}.  We also introduce the state-of-the-art velocity-based partitioning techniques in this section.

\subsection{TPR-tree and Rum-tree}
Saltenis et al. \cite{DBLP:conf/sigmod/SaltenisJLL00} proposed the TPR-tree (short for Time-Parameterized R-tree) which augments the R$^\star$-tree with velocities to index moving objects with motion functions. Specifically, an object in the TPR-tree is indexed by its time-parametrized position with respect to its velocity vector. A node in the TPR-tree is represented by a \textit{minimum bounding rectangle} (MBR) and the velocity on each side of the MBR which bounds all moving objects contained in the corresponding MBR at any time in the future. The TPR-tree uses time-parameterized metrics when choosing the target nodes for insertion and deletion. The time-parameterized metric is calculated as $\int_{t_l}^{t_l+H}A(t)dt$, where $A(t)$ is the metric used in the original R-trees. $H$ is the \textit{horizon} (the lifetime of the node) and $t_l$ is the time of an insertion or the index creation time. 


The TPR-tree uses a step-wise greedy strategy to choose the MBR where a new object is inserted. Since the objects are moving as time passes, the overlaps between MBRs become larger, which eventually makes the step-wise greedy strategy ineffective. Tao et al. proposed the TPR$^\star$-tree \cite{DBLP:conf/vldb/PapadiasTS03} that uses the same data structure as the TPR-tree with optimized insertion and deletion operations, which significantly reduce the overlaps between MBRs. 

Silva et al. proposed the Rum-tree \cite{DBLP:journals/vldb/SilvaXA09}, a variant of R-tree, which aims to reduce the cost of object updates through the so called \textit{update memo}. The RUM-tree processes updates through the update memo in main memory that avoids disk accesses for deleting old entries during an update process. The old entries are maintained by the \textit{garbage cleaner} inside the RUM-tree and are deleted lazily in batch mode. Therefore, the cost of an update operation in the RUM-tree is reduced to the cost of only an insert operation.


%

\subsection{B$^x$-tree and B$^{\mbox{dual}}$-tree}
The B$^x$-tree, proposed by Jensen et al. \cite{DBLP:conf/vldb/JensenLO04}, is the first indexing approach based on B$^+$-tree. The B$^x$-tree uses space-filling curves, such as Z-curves and Hilbert curves, to map the $d$-dimensional locations into scalars that can be indexed by B$^+$-trees. The time axis is partitioned into intervals of duration $\Delta t_{mu}$, which is the maximum duration in-between two updates of any object location. Each such interval is further partitioned into $n$ equal-length \textit{phases} and each phase is associated with a \textit{label timestamp}. Instead of indexing the object locations at their update timestamps, the B$^x$-tree indexes the locations at the nearest future label timestamp. After each $\Delta t_{mu}/n$ timestamps, one phase expires and another is generated. This rotation mechanism is essential to preserve the location proximity of the objects.

Yiu et al. \cite{DBLP:journals/vldb/YiuTM08} proposed the B$^{dual}$-tree that indexes the moving objects in the $2d$-dimensional \textit{dual space}, where velocity is considered as additional dimensions other than the $d$-dimensional location. The B$^{dual}$-tree applies a $2d$-dimensional Hilbert curve to map the underlining dual space to scalars and then indexes the scalars with B$^+$-trees. 


\subsection{STRIPES}
The quad tree \cite{DBLP:journals/acta/FinkelB74} is a hierarchical space partitioning structure, which can be augmented for indexing moving object. Patel et al. \cite{DBLP:conf/sigmod/PatelCC04} proposed the STRIPES, which indexes predicted trajectories in the \textit{dual transformed space}. Trajectories for objects in $d$-dimensional space are treated as points in the 2$d$-dimensional dual transformed space. This dual transformed space is then indexed using a regular hierarchical grid decomposition indexing structure which essentially employs a disk-based PR bucket quad tree structure \cite{DBLP:journals/csur/Samet84}. 


\begin{figure*}[!tp]
\centering
\begin{minipage}[b]{0.22\textwidth}
\subfigure[Index node]{
\centering
\includegraphics[height=0.8in]{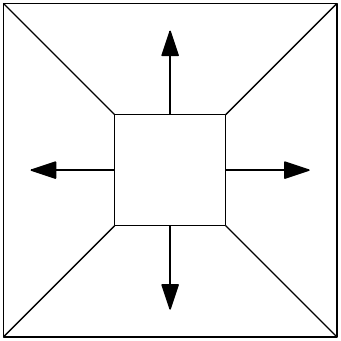}
\label{s1}
}
\end{minipage}
\begin{minipage}[b]{0.22\textwidth}
\subfigure[Search space expansion]{
\centering
\includegraphics[height=0.8in]{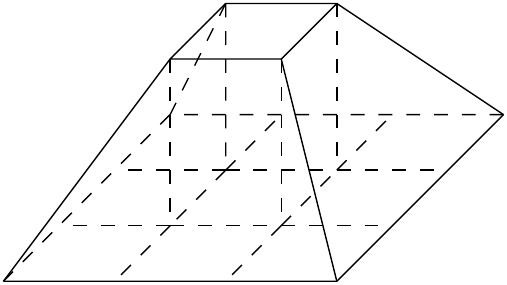}
\label{s11}
}
\end{minipage}
\caption{Search space expansion of an index node}
\label{space}
\end{figure*}

\subsection{Velocity-based partitioning}
Recently, velocity-based partitioning techniques, which utilize the velocity information from a global perspective, are used to further improve the query performance of indexes for MODs. Intuitively, velocity-based partitioning can improve query performance because search space expansion (defined as the enlargement of the index nodes) \cite{DBLP:journals/pvldb/NguyenHZW12} of the partitioned indexes considerably decreases in some scenarios.

Zhang et.al. \cite{DBLP:journals/pvldb/ZhangCJOZ09} firstly defined the VMBRs which represent the minimal rectangles in the velocity domain that bound the velocity vectors of all moving objects and proposed the partitioning method that minimizes the VMBRs within each partition. At the first step of this method, given the number of partitions $k$, the velocity vectors of exactly $k$ moving objects that form largest VMBR are selected as seeds for the $k$ partitions. Then each object is assigned to the partition with minimum VMBR increase. This method has some limitations. Firstly, it is difficult to determine the number of partitions $k$. Secondly, the partitioning might be far from optimum since this method relies on very simple heuristics and does not perform any analysis on search space expansion.

Thi et al. \cite{DBLP:journals/pvldb/NguyenHZW12} proposed the partitioning technique based on DVAs in the velocity domain. They applied principal component analysis and $K$-means clustering on the velocities of the moving objects to find $k$-1 DVAs. Then the velocity domain is partitioned into $k$ partitions according to the DVAs, one partition for each DVA plus one outlier partition. Each moving object is assigned to the nearest DVA partition if the distance between its velocity vector and the DVA is smaller than a threshold, otherwise it will be assigned to the outlier partition. Through this partitioning method, the velocity domain is reduced to nearly 1-dimensional parts, which dramatically reduces the search space expansion. However, this method still requires the number of partitions $k$ as a parameter. Moreover, the performance of this method will significantly reduce if the velocity domain has no effective DVAs.

In this paper, we propose a novel speed partitioning technique which dynamically and optimally partitions tree-based indexes based on speed values of the moving objects.

\section{The Optimization Problem}
\label{sec:prob}
In this section, we introduce the notion of search space expansion which can be used as a generic cost metric to estimate query performance of tree-based indexes for MODs. We then present the method for computing search space expansion and formulate the optimal speed partitioning problem.

\subsection{Search space expansion}
Figure \ref{s1} shows a typical example of how the geometry area of an index node expands. In this figure, the moving objects are originally located in a square area (the inner one) and move in arbitrary directions. At some future time, the objects will spread in a larger square area (the outer one). We model the expansion of the node as a trapezoid prism where the top base is the original area and the bottom base is the future area of the node. Figure \ref{s11} illustrates such a trapezoid prism of the node in Figure \ref{s1}. The volume of the trapezoid prism corresponding to an index node is called the search space expansion of this node. 
The sum of search space expansions of all index nodes is called the search space expansion of the index. A formal definition of search space expansion is given in Definition \ref{exp}.

\begin{definition}
  Search space expansion. Given any node in an MOD index $I$, its area at time $t$ is $S(t)$. The search space expansion of the node from time 0 to any future time $t_h$ is $\nu(t_h)=\int_0^{t_h}S(t)dt$. The search space expansion of the index is the sum of the search space expansions of all nodes: 
    $V(t_h)=\sum_{\forall node \in I}\nu(t_h)$
  \label{exp}
\end{definition}

If queries are randomly generated in the predefined space domain, nodes with larger search space expansions have higher probabilities to be accessed to answer the queries \cite{DBLP:conf/vldb/PapadiasTS03}. Consequently, indexes with smaller search space expansion enjoy better query performance. Thus we wish to find a partitioning strategy that minimizes the search space expansion of the indexes, i.e. the volumes of all trapezoidal prisms, in order to minimize query costs.

We propose the speed partitioning technique which partitions the indexes based on speed values of the moving objects. Since the moving objects are separated based on their speed values, thus fast growing nodes for high speed objects will not affect those for low speed objects. Therefore the search space expansion of an index will be dramatically reduced if we conduct appropriate partitioning on speed values. In the next subsection, we will discuss how to achieve the optimal index partitioning based on speed values. Note that in our analysis, we only consider the search space expansions of leaf nodes, because in most scenarios the number of leaf nodes significantly exceeds that of internal nodes.


\subsection{The optimal speed partitioning}
Our speed partitioning technique is based on solving the optimal speed partitioning problem, thus is different from and more generic than all state-of-the-art velocity-based partitioning techniques \cite{DBLP:journals/pvldb/ZhangCJOZ09} \cite{DBLP:journals/pvldb/NguyenHZW12} that rely on some kinds of heuristics. We now formalize the optimal speed partitioning problem that minimizes search space expansion. 

Denote $\mathcal{O}=\{o_1,o_2,\cdots, o_N\}$ as the set of moving objects and denote the speed of object $o_l$ as $v_{o_l}$. Let $\Omega=\{v_1, v_2, \dots, v_q\}$ represent the speed domain, where $v_1<v_2<\dots<v_q$. Thus for all $o_l \in \mathcal{O}$, we have $v_{o_l} \in \Omega$. We note that in most applications the speed domain can be easily discretized into finite number of different speed values. Let $v_0=v_1-\epsilon$, where $\epsilon$ is a positive number and $\epsilon\rightarrow 0$. $v_0$ is a dummy speed used for simplifying notations. Let $\Omega^+=\Omega \bigcup \{v_0\}$.

Now let $\Delta=\{\delta_0, \delta_1, \cdots, \delta_k\}$, $1\leq \delta_i\leq q$, where $\delta_0=0$ and $\delta_k=q$. Therefore $\Delta$ partitions the speed domain into $k$ (non-overlapping) parts, denoted as $\Omega_i=(v_{\delta_{i-1}},v_{\delta_{i}}]$, $1\leq i \leq k$. We say $\Delta$ is a partitioning on $\Omega$. Meanwhile, $\mathcal{O}$ is partitioned accordingly into $k$ parts: $P_i(1\leq i\leq k)$, where $P_i=\{o_l:v_{o_l}\in(v_{\delta_{i-1}},v_{\delta_{i}}]\}$. We denote $I_i$ as the corresponding indexing tree, such as the B$^x$-tree or the TPR$^\star$-tree, for $P_i$. Note that $k$ is automatically computed rather than an input of our method. 

Our goal is to find the optimal partitioning, denoted as $\Delta^\star$, that minimizes the overall search space expansion of all index partitions. We can achieve this goal by solving the following minimization problem:
\begin{equation}
 \Delta^\star = \underset{\Delta}{\arg\min}\left\{v_{\delta_0}< v_{\delta_1}< \cdots < v_{\delta_k}:V(t_h)\right\}
\label{opt}
\end{equation}
where $V(t_h)=\sum_{0<i\leq k}V_i(t_h)$ represents the overall search space expansion of all index partitions and $V_i(t_h)$ the search space expansion of partition $I_i$. $t_h$ is the maximum predict time for the predictive queries \cite{DBLP:conf/sigmod/SaltenisJLL00} \cite{DBLP:conf/vldb/PapadiasTS03}. Without loss of generality, we present next how to compute $V_i(t_h)$.


According to Definition \ref{exp}, in order to compute $V_i(t_h)$, we first need to compute the search space expansion of every single index node in $I_i$ which requires 1) the initial node area, and 2) the expanding speed of each node. We present the approach to compute $V_i(t_h)$ step by step in the following paragraphs.

\subsubsection{Generate uniform regions}
In most real world applications, the moving objects may not be uniformly distributed. Thus before calculating the search space expansion, we first divide the space domain into subregions such that the moving objects in $P_i$ are (close to) uniformly distributed within each subregion. Uniformity will not only significantly reduce the complexity of calculation but also help obtain more accurate estimations. We will introduce a quad tree based method to find the uniform subregions in Section \ref{sec:part}.
We denote the set of uniform subregions of $P_i$ as $\mathcal{R}_i=\{R_{i1}, R_{i2}, \dots, R_{i{m_i}}\}$.

\subsubsection{Compute initial node area}
Now we compute the initial areas of the nodes within subregion $R_{ij}$, where $0< j \leq m_i$. Without loss of generality, we assume $R_{ij}$ to be a square area with side length of $D_{ij}$. We also consider the index nodes as square shaped with expected side length of $d_{ij}$ and let $c$ represent the expected number of objects in each node. $c$ is determined by the storage size of each node which is a parameter in our method. Since moving objects are uniformly distributed in $R_{ij}$, we have $\frac{d_{ij}^2}{c}\propto \frac{D_{ij}^2}{N_{ij}}$ where $N_{ij}$ represents the number of objects in $R_{ij}$. Thus $d_{ij}$ can be estimated as
$d_{ij}=D_{ij}\sqrt{\frac{c}{N_{ij}}}$.

\subsubsection{Compute expanding speed}
Next we introduce the method for estimating expanding speeds of the index nodes in $R_{ij}$. Since we make no assumptions on the patterns of the moving objects' directions, we consider that the objects in each node travel at arbitrary directions. Thus every single node expands with equal speed in all directions while the expanding speed is the maximum speed value of the moving objects in the corresponding node.

Let $H_{iju}$ represent the number of moving objects in $R_{ij}$ whose speed values fall in the range $(v_{\delta_{i-1}}, v_u]$, where $v_u\in \Omega$ and $\delta_{i-1}< u\leq \delta_{i}$, formally
\begin{equation}
H_{iju}=|o_l\in R_{ij}:v_{\delta{i-1}}< v_{o_l} \leq v_u|
\label{defh}
\end{equation}
Since the speed values of the moving objects are independent given a certain speed distribution, expanding speed of any node in $R_{ij}$ is $v_u$ with the probability
\begin{equation}
p(i,j,u)=\frac{\binom{H_{iju}-H_{ij\delta_{i-1}}}{c}-\binom{H_{ij(u-1)}-H_{ij\delta_{i-1}}}{c}}{\binom{N_{ij}}{c}}
\end{equation}
where $\binom{a}{b}$ is a combination number. 

\subsubsection{Compute search space expansion}
\begin{figure}[!tp]
	\centering
	\begin{minipage}{.3\textwidth}
		\centering
		\includegraphics[height=1in]{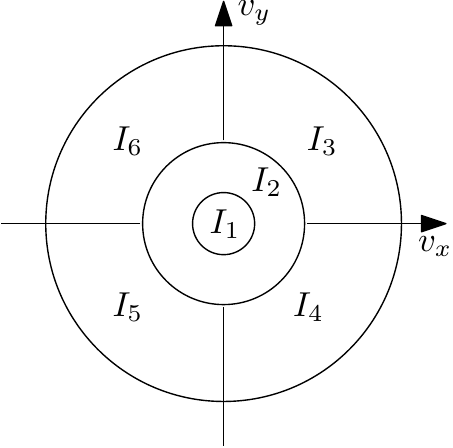}
		\caption{Speed partitioning}
		\label{spt}
	\end{minipage}%
	\begin{minipage}{0.7\textwidth}
		\centering
		\includegraphics[height=1in]{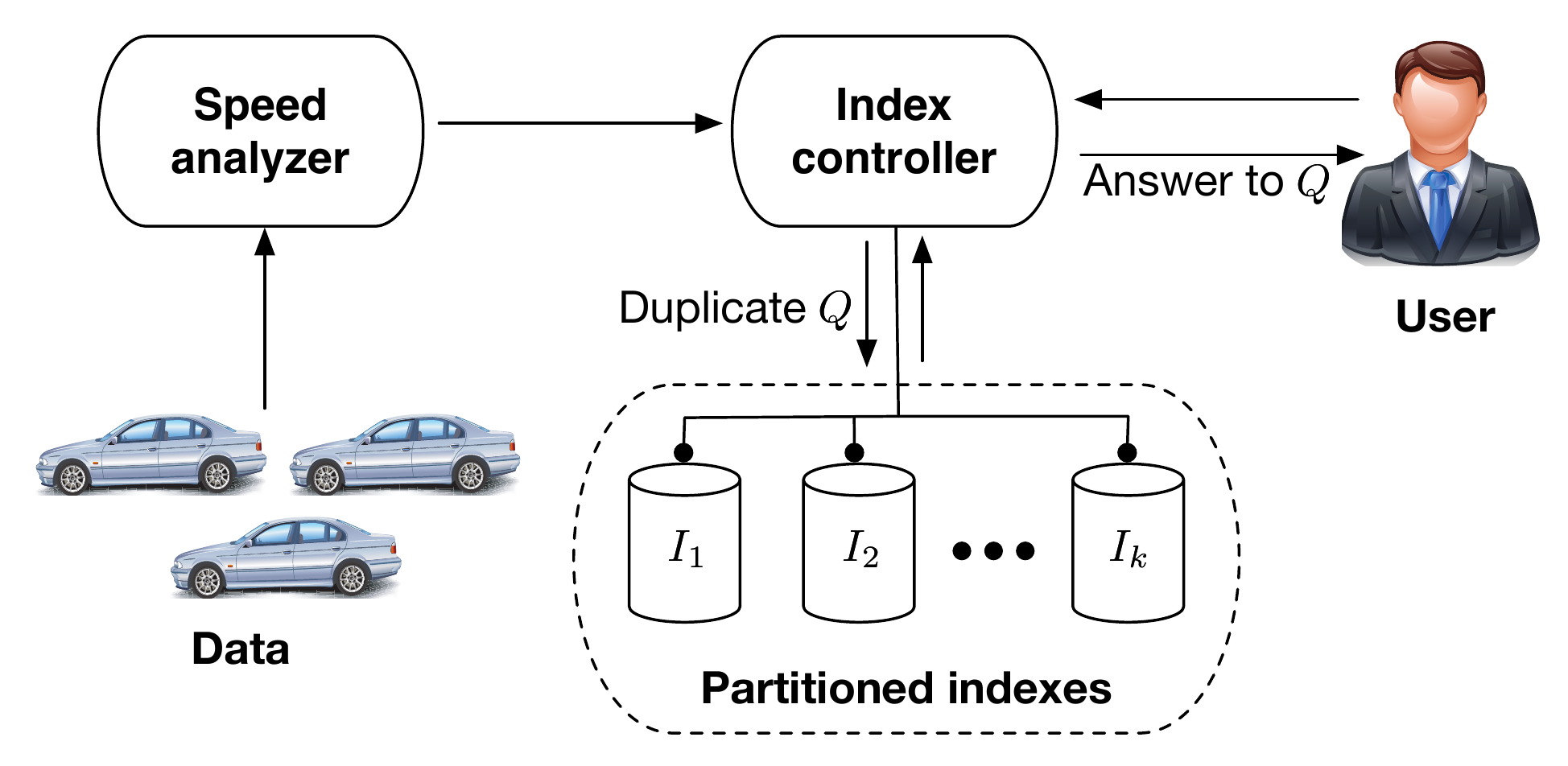}
		\caption{System architecture of SP}
		\label{model}
	\end{minipage}
\end{figure}

For each speed partition, we can apply a second-level (direction-based) partitioning into 4 quadrants as illustrated in Figure \ref{spt} if it further improves search space expansion. Hence we compute the search space expansion of $R_{ij}$ both with and without the second-level partitioning and select whichever achieves smaller value. When no second-level partitioning is performed, the search space expansion of a single node in $R_{ij}$ can be calculated by
\begin{equation}
\nu^1(t_h)=\nu(t_h,v_u)=\int_{0}^{t_h} \left(d_{ij}+2v_ut \right)^2 dt
\label{eq13}
\end{equation}
When the second-level partitioning is further applied, we compute the search space expansion for each quadrant. Expected side length of the nodes in the quadrant partitions is 2$d$ and the search space expansion is calculated by
\begin{equation}
\nu^2(t_h)=\nu(t_h,v_u)=\int_{0}^{t_h} \left(2d_{ij}+v_ut \right)^2 dt
\label{eq131}
\end{equation}
Therefore, the expected search space expansion of all nodes in $R_{ij}$ can be calculated by
\begin{equation}
  V_{ij}(t_h)=\left[ \frac{N_{ij}}{c} \right]\sum_{v_u\in \Omega, \delta_{i-1}<u \leq \delta_{i}} \nu(t_h) p(i,j,u)
\end{equation}
where $\left[ \frac{N_{ij}}{c} \right]$ computes the total number of nodes in $R_{ij}$ and $\nu(t_h)$ represents the minimum of $\nu^1(t_h)$ and $\nu^2(t_h)$. 
Finally, the overall search space expansion $V(t_h)$ is calculated by

\begin{equation}
  V(t_h)=\sum_{1\leq j\leq k} \sum_{0<j\leq m_i}V_{ij}(t_h)
  \label{overall}
\end{equation}


\section{The Partitioned Indexing System}
\label{sec:part}
Based on the above analysis on search space expansion, we propose the speed partitioning technique (SP) for indexing moving objects. Figure \ref{model} illustrates the system architecture of SP. SP uses a centralized indexing system consisting of three parts: the speed analyzer, the index controller, and the partitioned indexes. The speed analyzer receives data from the moving objects and computes the optimal speed partitioning. The index controller then creates the corresponding partitioned indexes. Once receiving queries from users, the index controller duplicates the queries and push them to the index partitions. After all index partitions finish processing the queries, the index controller collects and integrates the query results and sends them back to users. We will discuss more details of SP in the remainder of this section.

\subsection{The optimal speed partitioning}
In this subsection, we discuss how to find the optimal speed partitioning through dynamic programming. 

Let $\Lambda^\star_{r}$, $0<r\leq q$, be a sequence $(\lambda_0, \lambda_1, \cdots,\lambda_r)$ where $v_{\lambda_i}\in \Omega^\star$ and $0=\lambda_0<\lambda_1\leq \cdots\leq \lambda_{r-1}\leq \lambda_r=r$. The set of distinct values in $\Lambda^\star_{r}$ form the optimal partitioning of the sub speed domain of $(v_0,v_r]$, denoted as $\Delta^\star_{r}$. Thus our goal is to find $\Delta^\star_{q}$.

In order to compute $\Delta^\star_{q}$ using dynamic programming, we need to maintain two arrays $\mathbb{V}^\star$ and $\mathbb{T}^\star$, where $V^\star_{r}$ and $T^\star_{r}$ (the $r^{th}$ values of $\mathbb{V}^\star$ and $\mathbb{T}^\star$) store the search space expansion of $\Delta^\star_{r}$ and the $r^{th}$ value ($\lambda^\star_{r-1}$) in $\Lambda^\star_{r}$, respectively.
$V^\star_{r}$ and $T^\star_r$ can be computed by Equation (\ref{vstar}) and (\ref{tstar}), respectively.
\begin{equation}
  V^\star_{r} =
  \left\{
\begin{array}{ll}
0&r=0\\
\min\limits_{0\leq s< r}\{ V^\star_{s}+V_{(v_s,v_r]} \}&0<r\leq q
\end{array}
\right.
\label{vstar}
 \end{equation}
\begin{equation}
  T^\star_{r} = \underset{0\leq s< r}{\arg\min}\{V^\star_{s}+V_{(v_s,v_r]} \},0<r\leq q
  \label{tstar}
\end{equation}
where $V_{(v_s,v_r]}$ is the search space expansion of partition $P_{(v_s,v_r]}$ and $P_{(v_s,v_r]}=\{o_l:v_{o_l}\in(v_s,v_r]\}$. Note that we define $V^\star_0=0$ in order to simplify denotations. Next we discuss how to compute $V_{(v_s,v_r]}$, for all $(v_s,v_r]\subset \Omega$.

\begin{algorithm}[!tp] 
  \caption{$merge(\mathcal{Q})$}
  \label{algmerge}
  \SetKwInOut{Input}{input}\SetKwInOut{Output}{output}
   \Input{$\mathcal{Q}$: a set of quad tree nodes}
   \Output{$\mathcal{R}$: a set of uniform subregions}
   \tcc{check the uniformity of the current nodes}
   \eIf{$\forall Q_j\in \mathcal{Q}$, $Q_j$ is uniform}{
     add the region of $\mathcal{Q}$ to $\mathcal{R}$\;
   }
   {
   \tcc{explore the child nodes}
     \For{$i\leftarrow 0$ \KwTo $3$}{
       \ForEach{$Q_j \in \mathcal{Q}$}{
         $CQ_j\leftarrow Q_j.child[i]$\;
       }
       $merge(\mathcal{CQ})$\;
     }
   }
\end{algorithm}

\begin{figure}[!tp]
	\centering
	\includegraphics[height=1in]{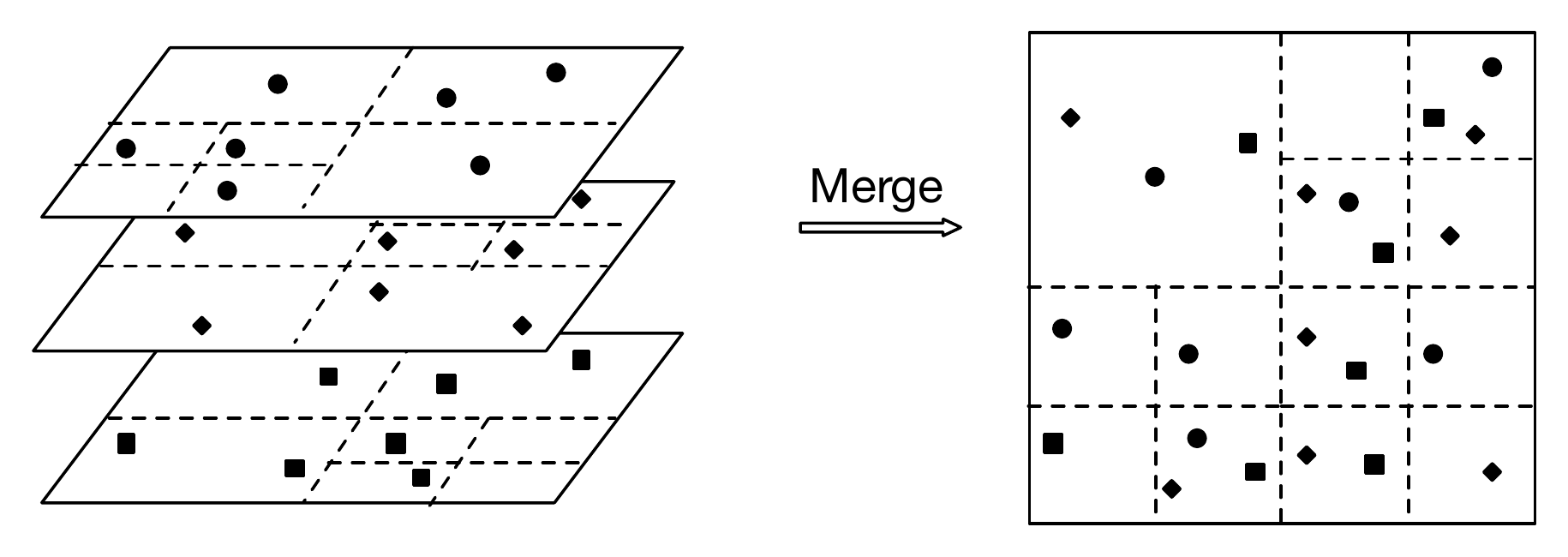}
	\caption{An example of merge}
	\label{merge}
\end{figure}

In order to compute $V_{(v_s,v_r]}$ using Equation (\ref{overall}), we first need to generate the uniform subregions mentioned in Section \ref{sec:prob}. We propose a quad tree \cite{DBLP:journals/acta/FinkelB74} based method to generate the uniform subregions for every $P_{(v_s,v_r]}$. We first divide the objects into $q$ layers, where moving objects within the same layer have same speed values (represented by the average speed value in each layer). Each layer is divided into square subregions using a quad tree such that the objects in each subregion are uniformly distributed. We use $\chi^2$-test (significance level 5\%) to test the uniformity of each subregion. We also fix 5 as the maximum depth of the quad trees. In order to generate the uniform subregions for $P_{(v_s, v_r]}$, we need to combine the corresponding layers, layer $s+1$ through $r$. We choose the most fine grained division when the divisions of different layers conflict, thus objects in the subregions of the combined layer always contain uniformly distributed objects. Figure \ref{merge}(left) shows an example of such layers, where there are 3 different speed values $v_1$, $v_2$, $v_3$ and the objects in the 3 layers are represented as squares, diamonds, and dots, respectively. Figure \ref{merge}(right) shows the result of the merge operation. 

Algorithm \ref{algmerge} shows the pseudo code for the merge operation. This is a recursive algorithm which takes a set of $r-s$ quad tree nodes (one node for each layer) as input. If objects within all the current nodes are uniformly distributed, we add the (square) spatial region represented by the quad tree nodes into the result set (lines 1-2). Otherwise, we recursively explore the 4 child nodes (each 2-dimensional quad tree node has 4 child nodes) at the next level of the quad trees (lines 3-7). Note that the input nodes will always locate at the same positions in the corresponding quad trees for all recursive calls, since we set the root nodes of the quad trees as input of the initial call. 

\begin{algorithm}[!tp]
	Create quad trees\;
   \tcc{Pre-compute search space expansion for partition $P_{(v_s,v_r]}$ using Equation (\ref{overall})}
 \ForEach{$(v_s, v_r] \in \Omega$}{
    $V_{(v_s,v_r]}\leftarrow$ the search space expansion of $P_{(v_s,v_r]}$\;
  }
  \tcc{Iteratively compute $V^\star_{r}$ and $T^\star_{r}$ using Equation (\ref{vstar}) and (\ref{tstar})}
   $V^\star_{0}\leftarrow 0$\; 
    \For{$r\leftarrow 1$ \KwTo $q$}{
      $min\leftarrow inf$\;
      \For{$s\leftarrow 0$ \KwTo $r-1$}{
        \If{$V^\star_{s} + V_{(v_s,v_r]}<min$}{
          $min\leftarrow V^\star_{s} + V_{(v_s,v_r]}$\;
          $T^\star_{r}\leftarrow s$\;
      }
    }
      $V^\star_{r}\leftarrow min$\;
    }
  
  \tcc{Compute the final results using Equation (\ref{lamb})}
   $\lambda_q\leftarrow q$\;
     \For{$i\leftarrow q-1$ \KwTo 1}{
   $\lambda_i\leftarrow T^\star_{\lambda_{i+1}}$\;
   }
   $\lambda_0\leftarrow 0$\;
 \caption{Find the optimal speed partitioning}
 \label{algdp}
\end{algorithm}

In order to find the optimal partitioning $\Delta^\star_{q}$, we need to compute $V^\star_{r}$ for each $r$ ($0< r\leq q$). As shown in Equations (\ref{vstar}) and (\ref{tstar}), we iteratively find the best $s$ which leads to the optimal partitioning on $(v_0,v_r]$ and stores it as $T^\star_{r}$. During the computation for $V^\star_{r}$, we can use previously computed optimal results on $(v_0, v_s]$, i.e. the values of $V^\star_{s}$ for each $s$ ($0\leq s< r$). Finally, we can obtain the optimal partitioning on $(v_0,v_q]$ by tracking backwards the values in $\mathbb{T}^\star$, i.e. each $\lambda_i\in \Lambda^\star_{q}$ ($0\leq i\leq q$) can be computed by
\begin{equation}
  \lambda_i=
\left\{
\begin{array}{ll}
  0&i=0\\
  T^\star_{\lambda_{i+1}}&0< i<q\\
  q&i=q\\
\end{array}
\right.
\label{lamb}
\end{equation}

Algorithm \ref{algdp} shows the pseudo code of our dynamic programming based algorithm to solve the optimal speed partitioning problem. Algorithm \ref{algdp} first creates the quad trees for uniform subregion generation (line 1). Then the search space expansions of partition $P_{(v_s,v_r]}$, for all $(v_s,v_r]$, are calculated (lines 2-3). Then dynamic programming is used to compute the values of $V^\star_{r}$ and $T^\star_{r}$ based on Equations (\ref{vstar}) and (\ref{tstar}) (lines 4-11). Finally, $\lambda_0$ through $\lambda_q$ are computed from $\mathbb{T}^\star$ using Equation (\ref{lamb}) (lines 12-15). 
Note that we compute the search space expansion (line 3) both with and without the second-level partitioning as described in Section \ref{sec:prob} and store the smaller value as $V_{(v_s,v_r]}$. The corresponding speed partition in the final result is further partitioned into four sub-partitions (one for each quadrant in the velocity domain) if it achieves smaller search space expansion. Figure \ref{spt} shows an example of the output of our algorithm. Actually, high speed partitions are more likely to be further partitioned into quadrants since direction has more impact on high speed partitions. The time complexity of Algorithm \ref{algdp} is analyzed as follows.

\subsubsection{Complexity analysis}
Execution time of Algorithm \ref{algdp} consists of three parts: 1) creating the quad trees takes $O(N)$ time; 2) pre-computing the search space expansions for each sub speed domain takes $O(q^2)$ time; and 3) the dynamic programming part also takes $O(q^2)$ time. Thus the total time complexity of Algorithm \ref{algdp} is $O(N + q^2)$. Note that the analysis relies on the condition that maximum depth of the quad trees is fixed, as mentioned earlier in this section.

\subsection{Index update}
Index update of our system consists of two parts: object update and partition update. Object update corresponds to status (e.g. location and velocity) updates of the moving objects, which is essential to keep the objects' locations up-to-date. When a moving object updates its status, the index controller will determine whether it should be inserted into a different partition based on its current velocity. Then the object will be either deleted from its previous partition and inserted into the new one or simply updated in the previous partition. Note that each index partition contains only a portion of the moving objects, thus object update in the partitioned indexes takes less CPU time than that in the original index without partitioning.

Partition update corresponds to changes of the optimal speed partitioning. Since the objects are continuously moving, both their location and speed distributions might change over time. Thus we need to re-compute the uniform subregions as well as the optimal speed partitioning when necessary. We simply conduct partition updates periodically with cycle time customized according to the data set. For example, in city road networks, location and speed distributions of the vehicles might be different between rush hours and regular hours, for which we can use hourly partition update routines.

\subsection{Query processing}
In this work, we evaluate predictive time-slice queries \cite{DBLP:conf/sigmod/SaltenisJLL00} \cite{DBLP:conf/vldb/PapadiasTS03} which retrieve tentative future locations of the moving objects. We consider both predictive range queries and predictive $k$ nearest neighbor ($k$NN) queries. A predictive range query is associated with two coordinates (bottom-left point and upper-right point of the range query window) and the predict time, while a predictive $k$NN query is associated with a coordinate (center of the $k$NN query), $k$NN-$k$, and the predict time. 

Query processing for SP is straightforward. The original queries are duplicated (with modifications if necessary) and processed within each partition either concurrently or sequentially. In order to compare the performance between partitioned indexes and their unpartitioned counterparts, in this paper, we conduct the duplicated queries sequentially. Within each index partition, queries are performed using the algorithm associated with the corresponding indexing structure (e.g. the B$^x$-tree or the TPR$^\star$-tree).

\section{Experimental Study}
\label{sec:expe}
In this section, we conduct extensive experiments to evaluate the performance of our speed partitioning technique for both main memory indexes and external memory or disk indexes. Both simulated traffic data and real world GPS tracking data are used in the experiments. 
We evaluate both throughput and query response time. Query response time consists of I/O latency and CPU time for disk indexes while only CPU time for main memory indexes.

We use the B$^x$-tree and the TPR$^\star$-tree as the basic indexing structures. 
We compare our approach of speed partitioning (SP-B$^x$ and SP-TPR$^\star$) with the state-of-the-art approaches of DVA-based partitioning \cite{DBLP:journals/pvldb/NguyenHZW12} (dVP-B$^x$ and dVP-TPR$^\star$) and VMBR-based partitioning \cite{DBLP:journals/pvldb/ZhangCJOZ09} (mVP-B$^x$ and mVP-TPR$^\star$) as well as the baseline approaches (B$^x$ and TPR$^\star$). We set the number of partitions $k$ in DVA and VMBR-based partitioning techniques as 3 and 5, respectively, which is consistent with the experimental settings in the original papers.
All algorithms are implemented with C++ language and all experiments are performed with 2.93GHz Intel Xeon CPU and 1TB RAM in CentOS Linux . 

\begin{figure*}[!tp]
	\centering
	\subfigure[Part of Seoul road network]{
		\centering
		\includegraphics[width=0.22\textwidth, height=1in]{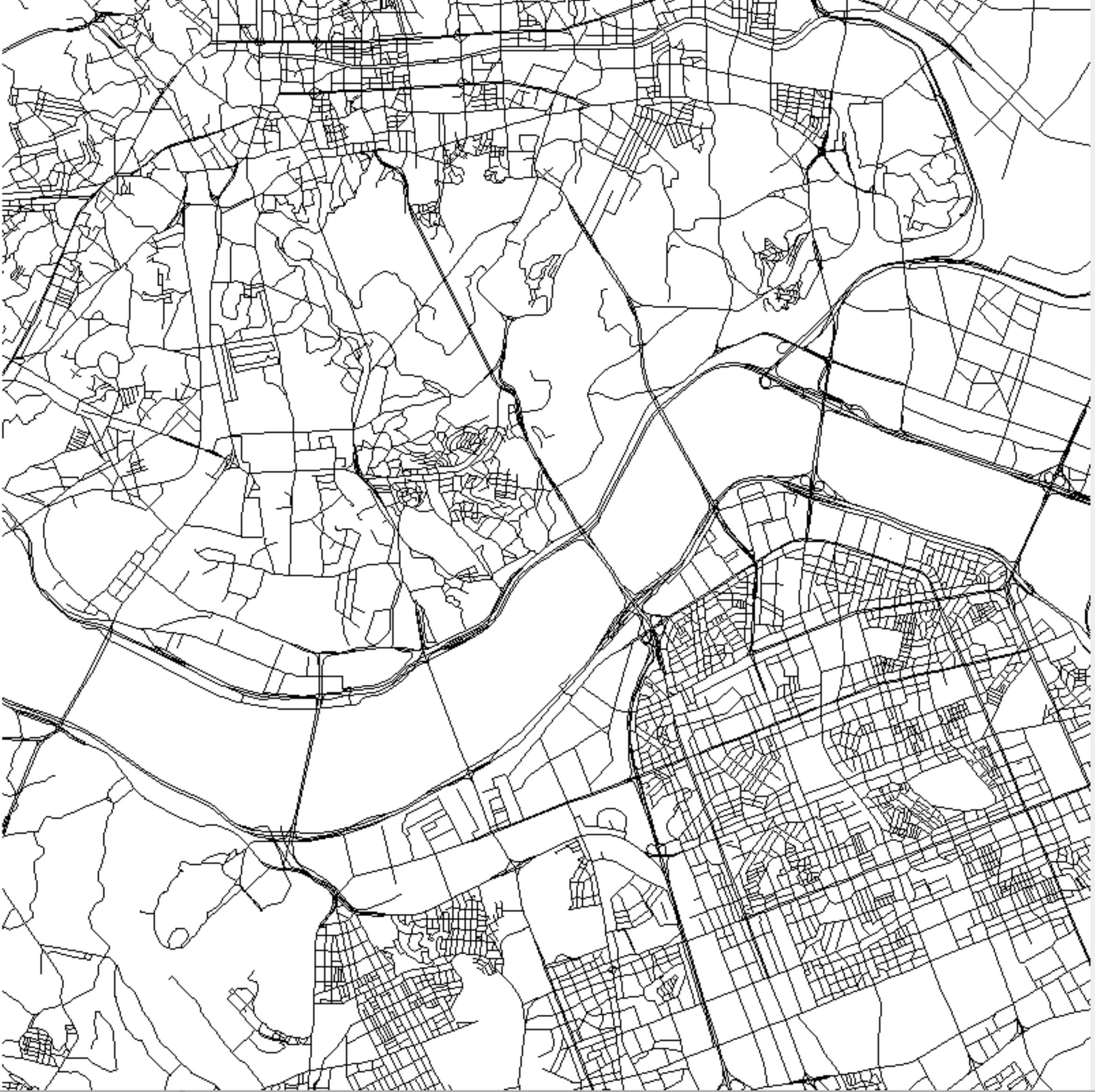}
		\label{seo}
	}
	\subfigure[Part of London road network]{
		\centering
		\includegraphics[width=0.22\textwidth, height=1in]{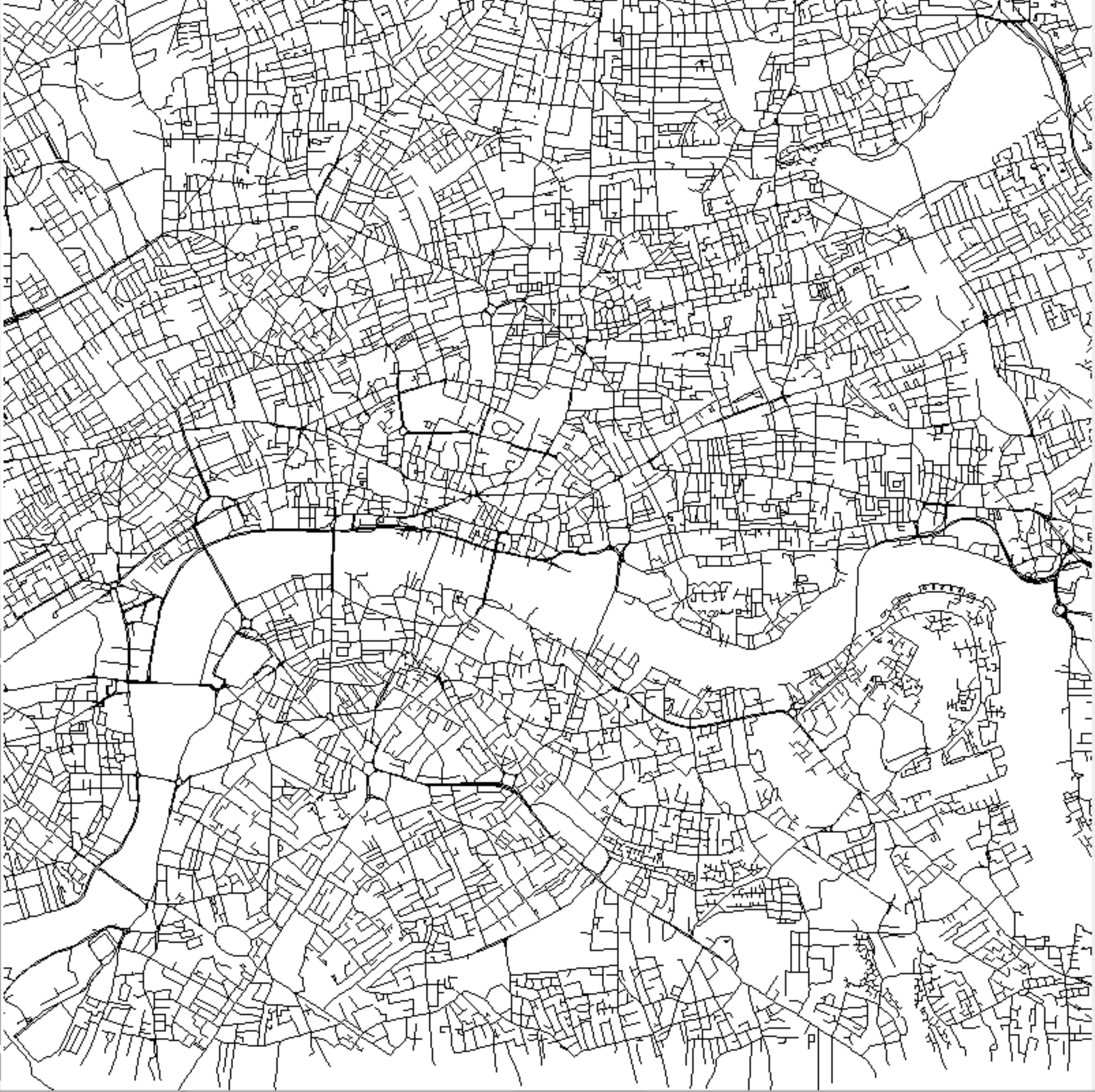}
		\label{ld}
	}
	\subfigure[Part of Boston road network]{
		\centering
		\includegraphics[width=0.22\textwidth, height=1in]{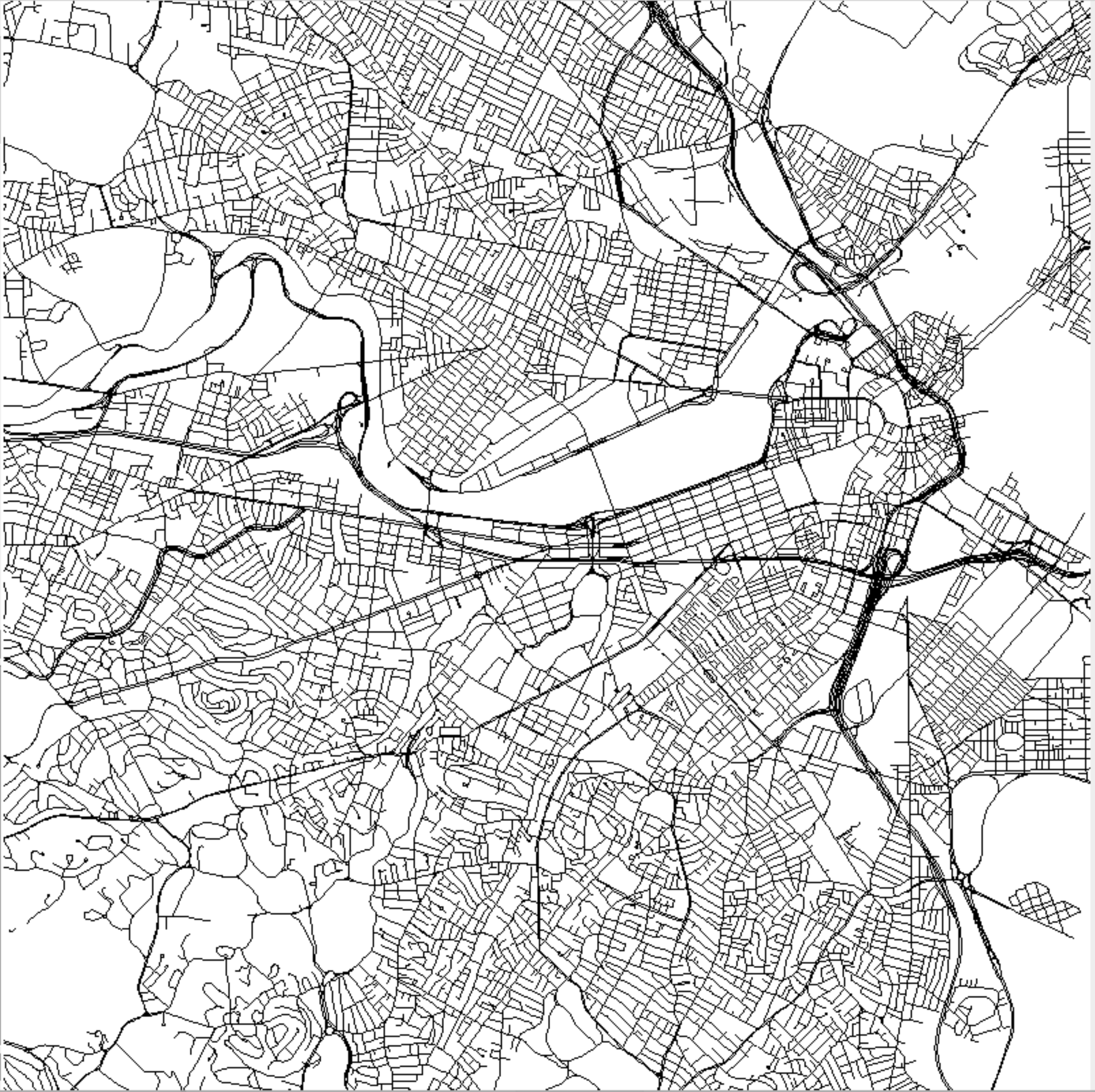}
		\label{bos}
	}
		\subfigure[Part of Shenzhen road network]{
			\centering
			\includegraphics[width=0.22\textwidth, height=1in]{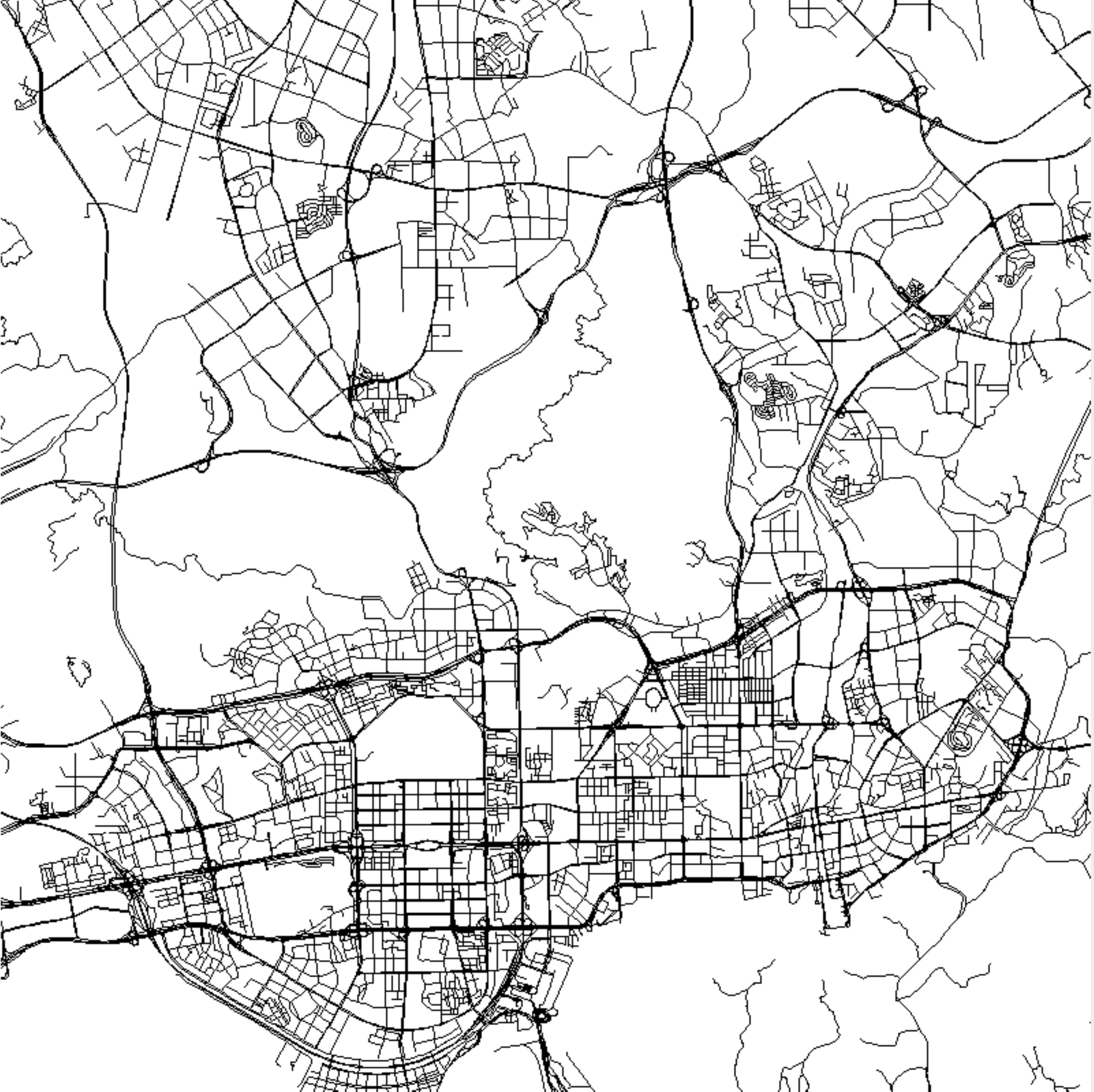}
			\label{sz}
		}
	\caption{City road networks for traffic simulation}
	\label{network}
\end{figure*}

\begin{table}[!tp]\scriptsize
\centering
\caption{Experimental settings}
\begin{tabular}{|l||l|} \hline
\textbf{Parameter} & \textbf{Setting}\\\hline
Space domain(m$\times$m) & \textbf{10,000$\times$10,000}\\\hline
Number of objects & \textbf{100K}, 200K, $\dots$, 500K\\ \hline
Query window size (m$\times$m)& 200$\times$200, \textbf{400$\times$400}, $\dots$, 1000$\times$1000\\\hline
$k$NN - $k$ & 10, 20, \textbf{30}, $\dots$, 50\\ \hline 
Query predict time (ts) & 0, 30, \textbf{60}, $\dots$,120\\\hline
Node size (byte) & 1K, 2K, \textbf{4K},$\cdots$, 16K\\\hline
Data sets & SEO, \textbf{LD}, BOS, SZ\\\hline
\end{tabular}
\label{setting}
\end{table}

\subsection{Data sets}
In this subsection, we introduce data sets used in the experiments. Figure \ref{network} shows city road networks corresponding to the data sets in the experiments. The experimental settings are displayed in Table \ref{setting} where the default settings are boldfaced.

\subsubsection{Simulated traffic data}
The simulation of city traffic consists of two parts: road network generation and traffic generation. City road networks are generated from the XML map data downloaded from http://www.openstreetmap.org. Our traffic generator is based on the digital representation of real road networks and the network-based moving object generator of Brinkhoff \cite{DBLP:journals/geoinformatica/Brinkhoff02}. A road is a polyline consisting of a sequence of connected line segments. The initial location of a moving object is randomly selected on the road segments. The object then moves along this segment in either direction until reaching crossroads, where it has a 25\% chance to stop for several seconds due to the traffic and then continues moving along another randomly selected connected segment. 

We assume speed values of the moving vehicles in each road segment follow a random variable $X$ and $X\sim \mathcal{N}(\mu,\sigma^2)$, where $\mathcal{N}$ is the normal distribution, $\mu$ and $\sigma$ are set according to categories of the road segments. We divide the road segments into three categories: C1) freeways/motorways with fastest traffic, C2) primary roads with secondary fastest traffic, and C3) street ways or residential roads with slowest traffic. We randomly select the normal distribution parameter $\mu$ from a range in terms of m/s for each category: C1) [25, 40], C2) [5, 25], C3) [0, 15]. We set $\sigma$=10 m/s for all road segments. We simulate the traffic in a time period of 120 seconds and assume that location and speed distributions of the moving objects remain unchanged during the monitored time period.  

\subsubsection{GPS tracking data}
The SZ data set contains 100K trajectories of taxis within the urban area of Shenzhen, China. Each trajectory contains a sequence of GPS tracking data with timestamps in a single day. The trajectories are not sampled with equal time intervals and the smallest sampling interval is 15 seconds. The data set can be accessed at http://mathcs.emory.edu/aims/spindex/taxi.dat.zip. 
%

\subsection{Experimental results}

\begin{figure}[!tp]
	\centering
	\subfigure[Varying $N$]{
		\centering
		\begin{minipage}[b]{0.22\textwidth}
			\includegraphics[width=1.1\textwidth,height=1in]{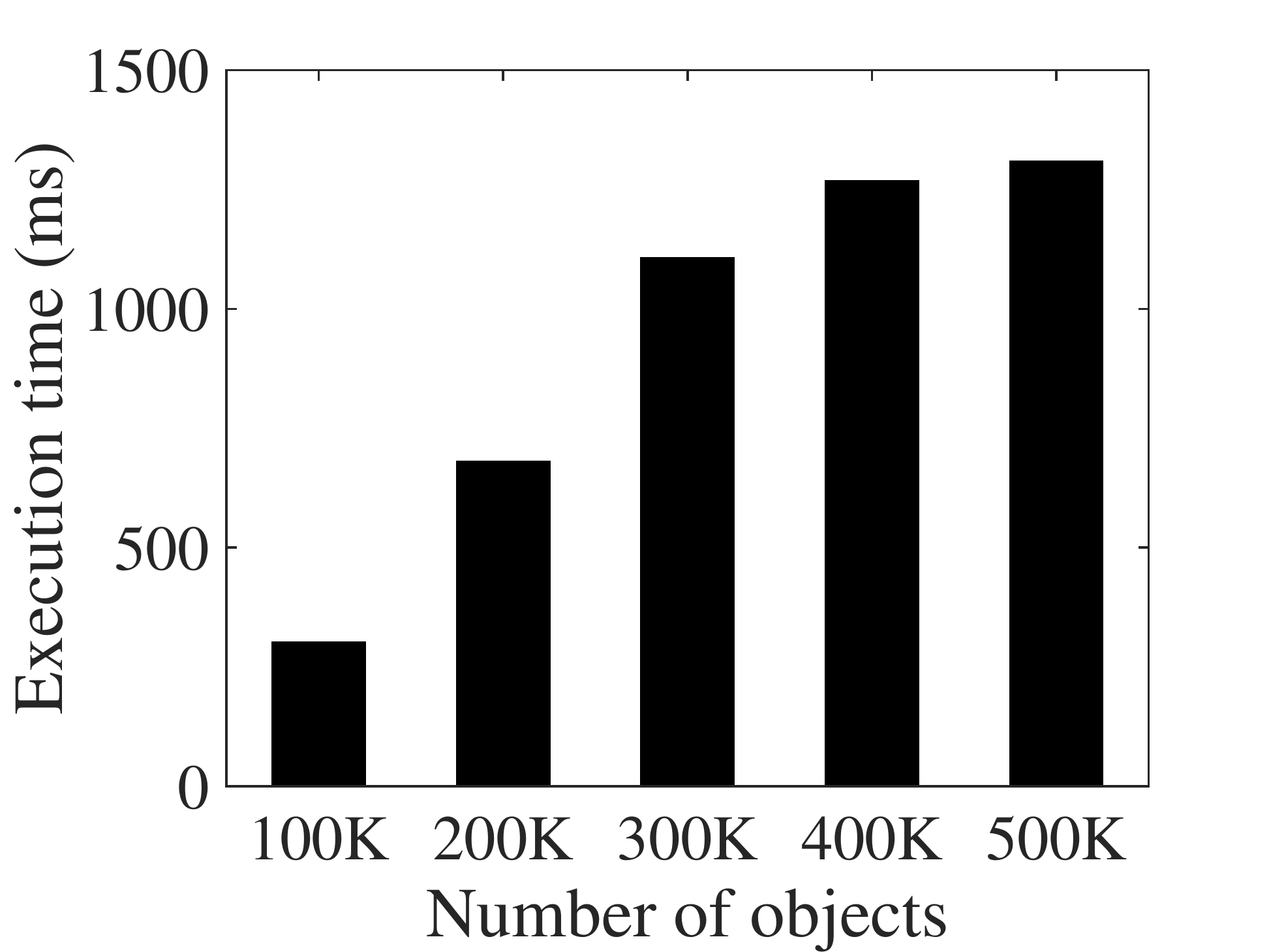}
		\end{minipage}
		\label{no_over}
	}
	\subfigure[Varying $q$]{
		\centering
		\begin{minipage}[b]{0.22\textwidth}
			\includegraphics[width=1.1\textwidth,height=1in]{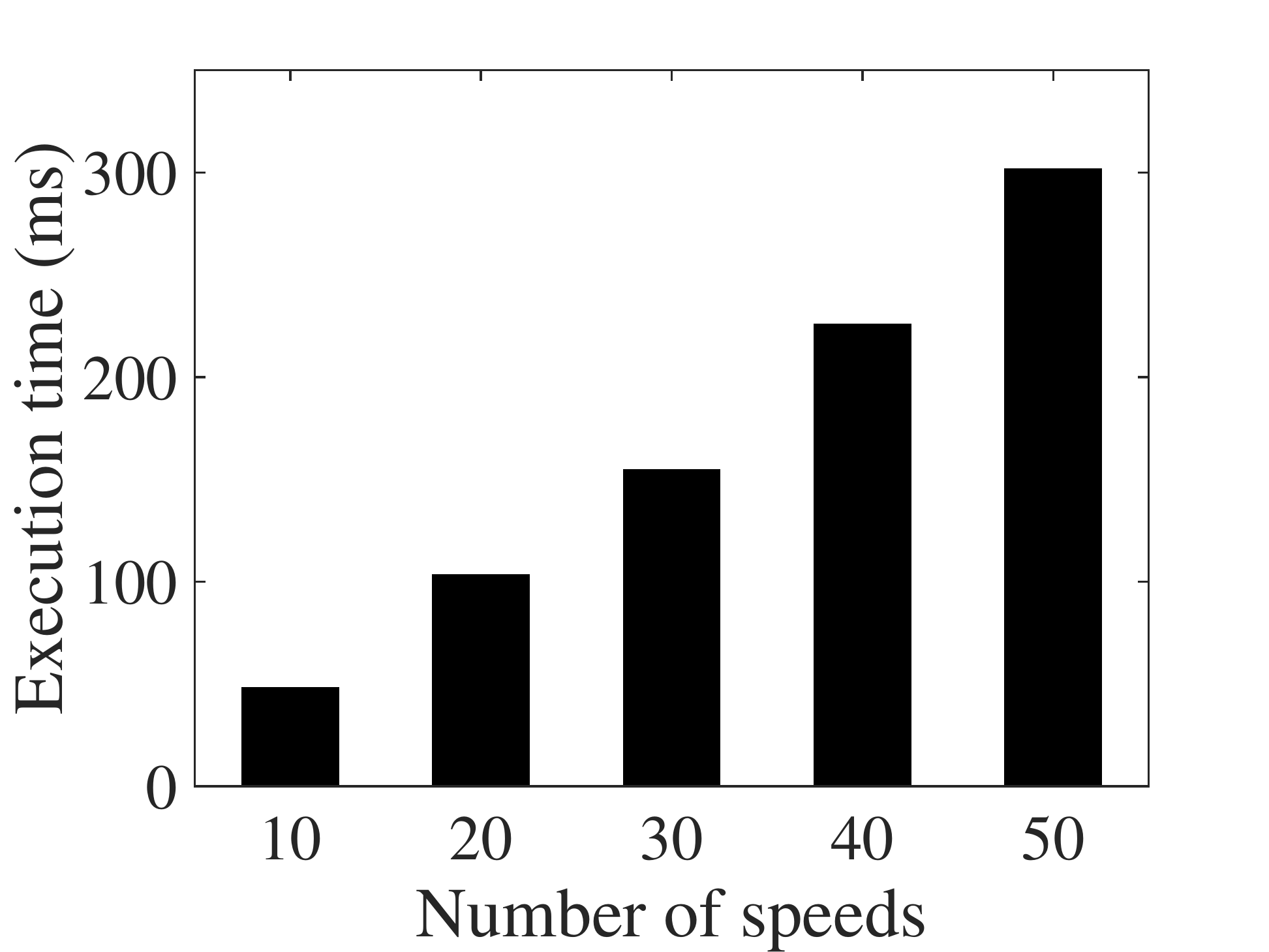}
		\end{minipage}
		\label{ns_over}
	}
	\caption{Execution time of SP}
	\label{exp-over}
\end{figure}

\begin{figure}[!tp]
	\centering
	\subfigure[Throughput]{
		\centering
		\begin{minipage}[b]{0.22\textwidth}
			\includegraphics[width=1.1\textwidth,height=1in]{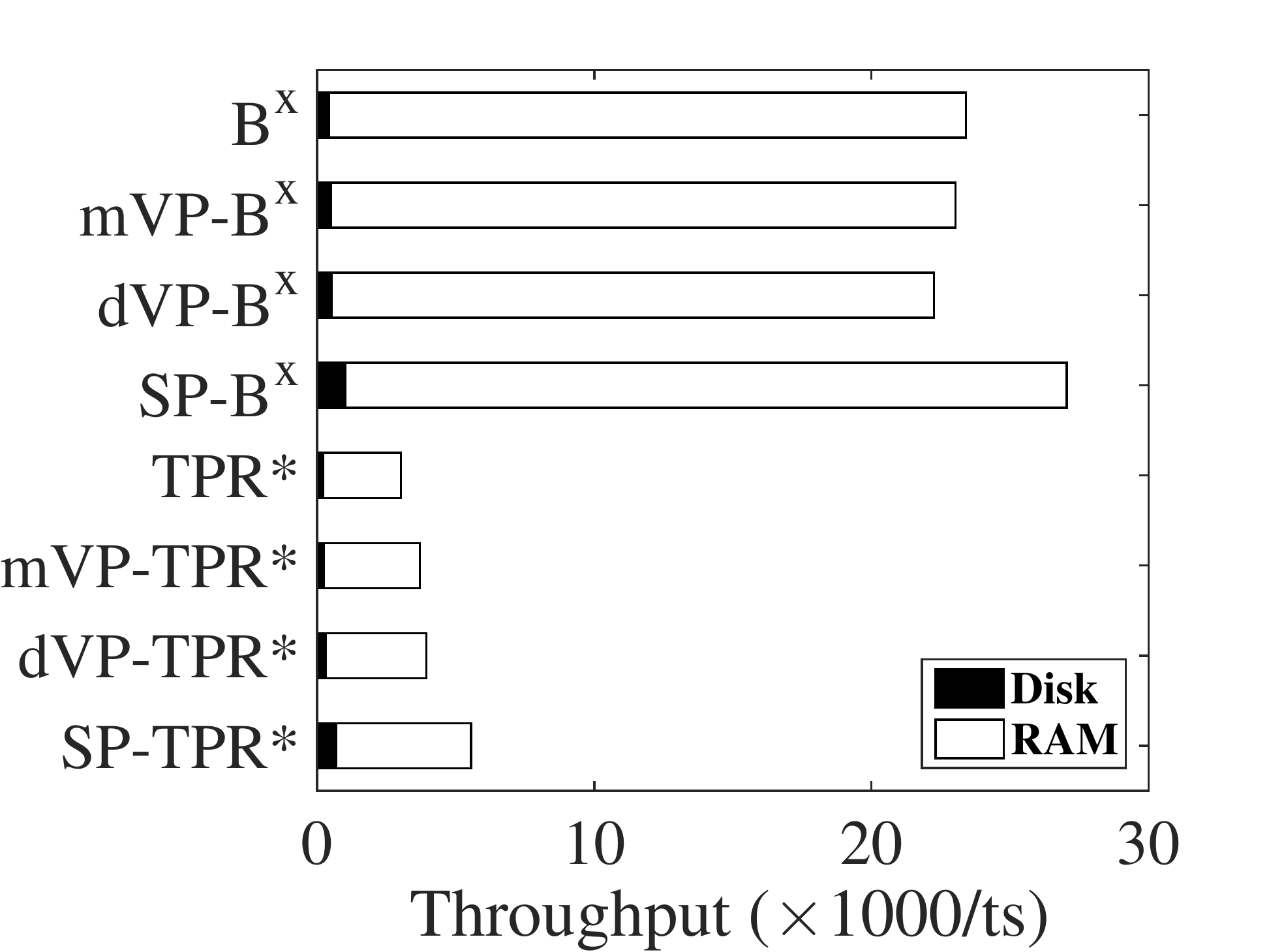}
		\end{minipage}
		\label{cs_th}
	}
	\subfigure[Range query]{
		\centering
		\begin{minipage}[b]{0.22\textwidth}
			\includegraphics[width=1.1\textwidth,height=1in]{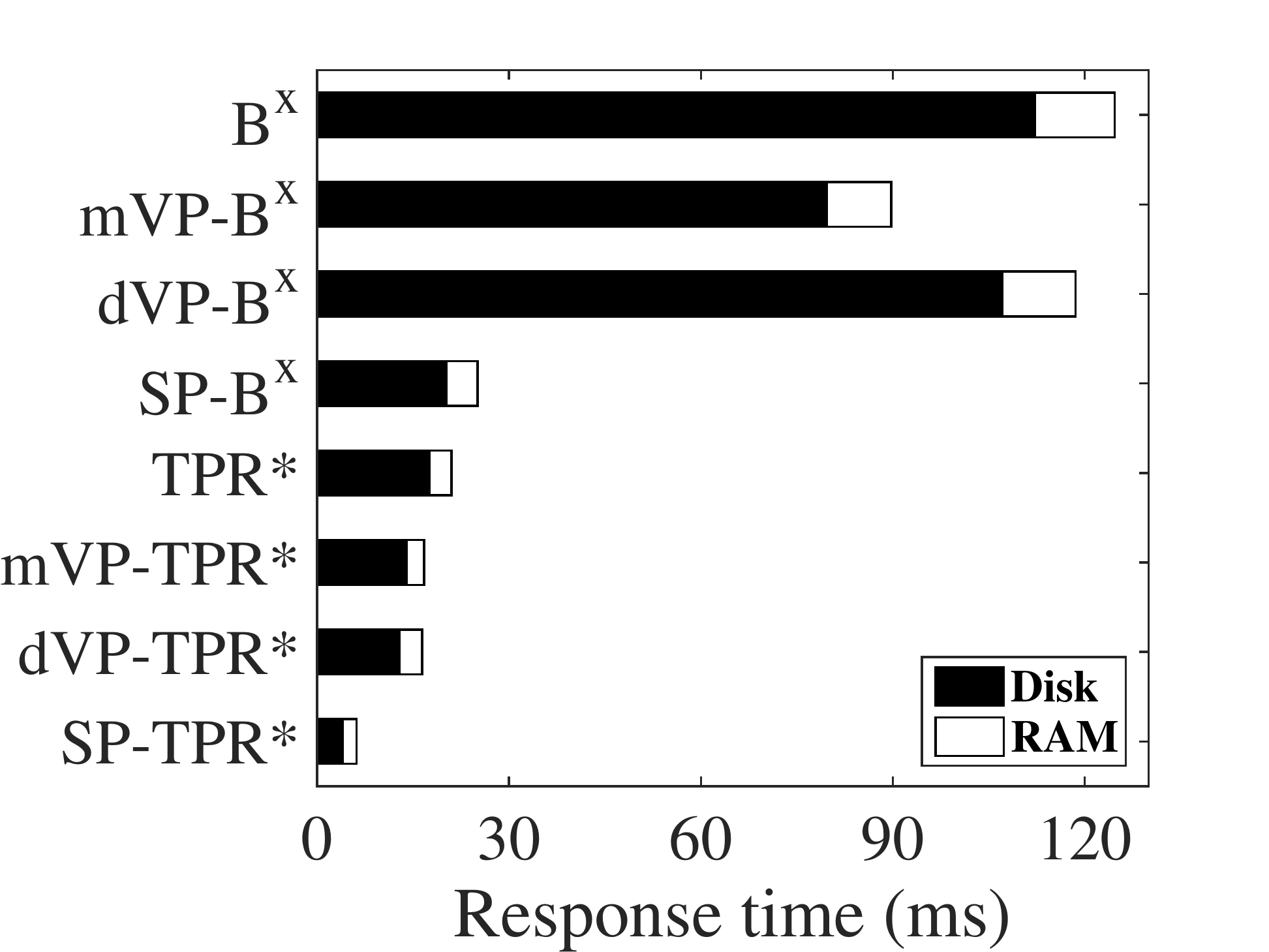}
		\end{minipage}
		\label{cs_range}
	}
	\subfigure[$k$NN query]{
		\centering
		\begin{minipage}[b]{0.22\textwidth}
			\includegraphics[width=1.1\textwidth,height=1in]{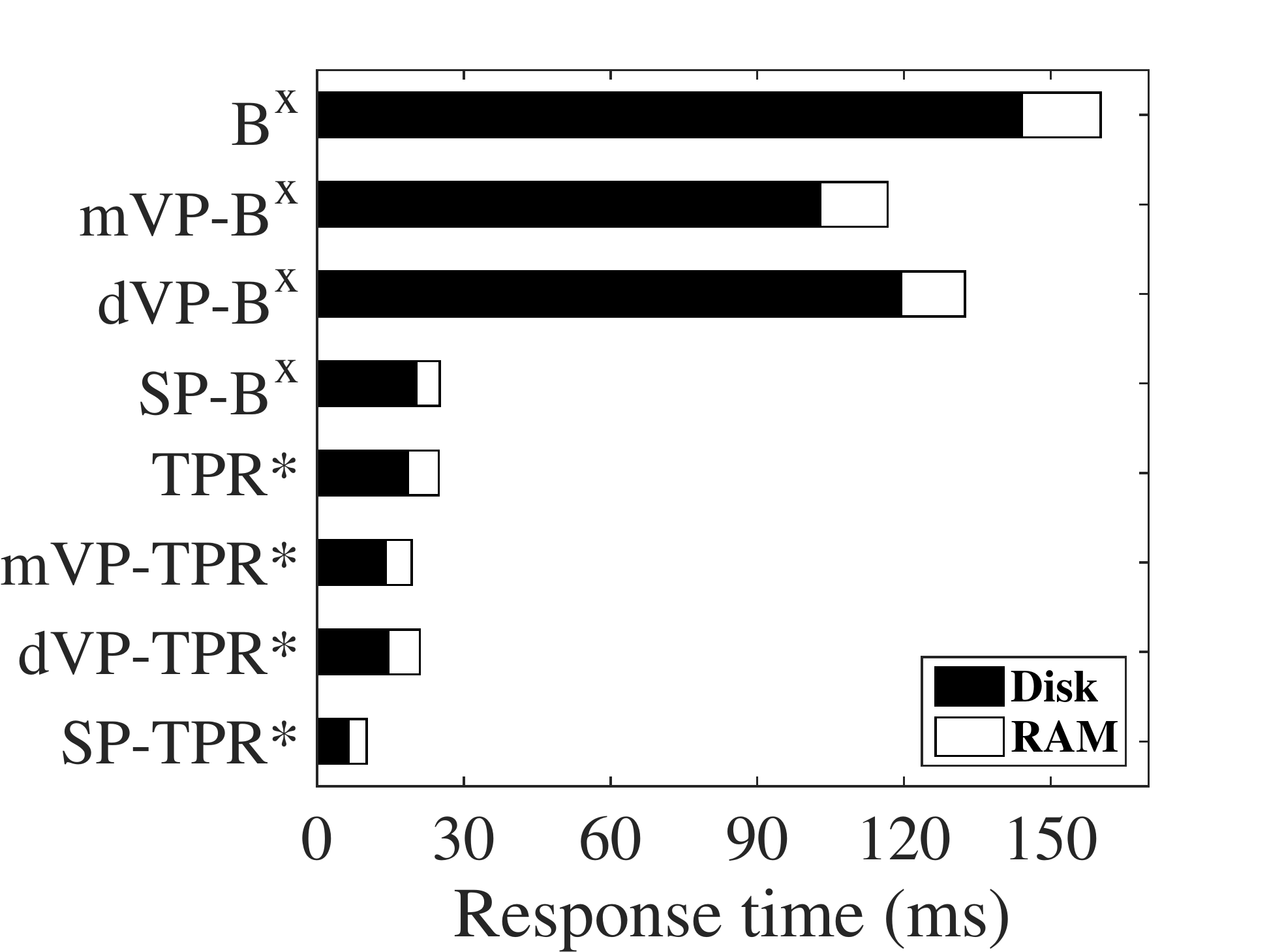}
		\end{minipage}
		\label{cs_knn}
	}
	\caption{Disk indexes and RAM indexes}
	\label{exp-cache}
\end{figure}

\begin{figure}[!tp]
	\centering
	\subfigure[Throughput]{
		\centering
		\begin{minipage}[b]{0.22\textwidth}
			\includegraphics[width=1.1\textwidth,height=1in]{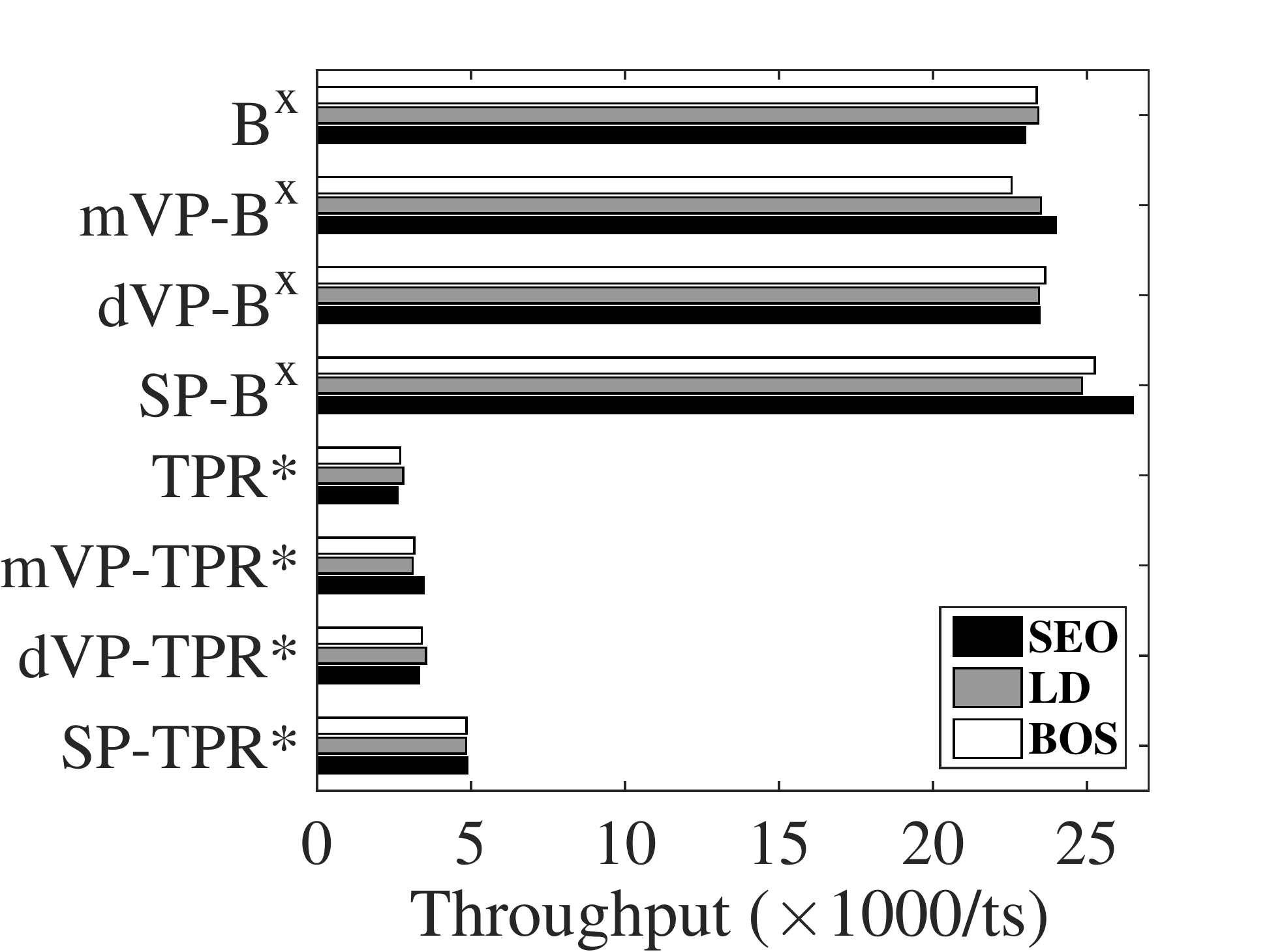}
		\end{minipage}
		\label{da_th}
	}
	\subfigure[Range query]{
		\centering
		\begin{minipage}[b]{0.22\textwidth}
			\includegraphics[width=1.1\textwidth,height=1in]{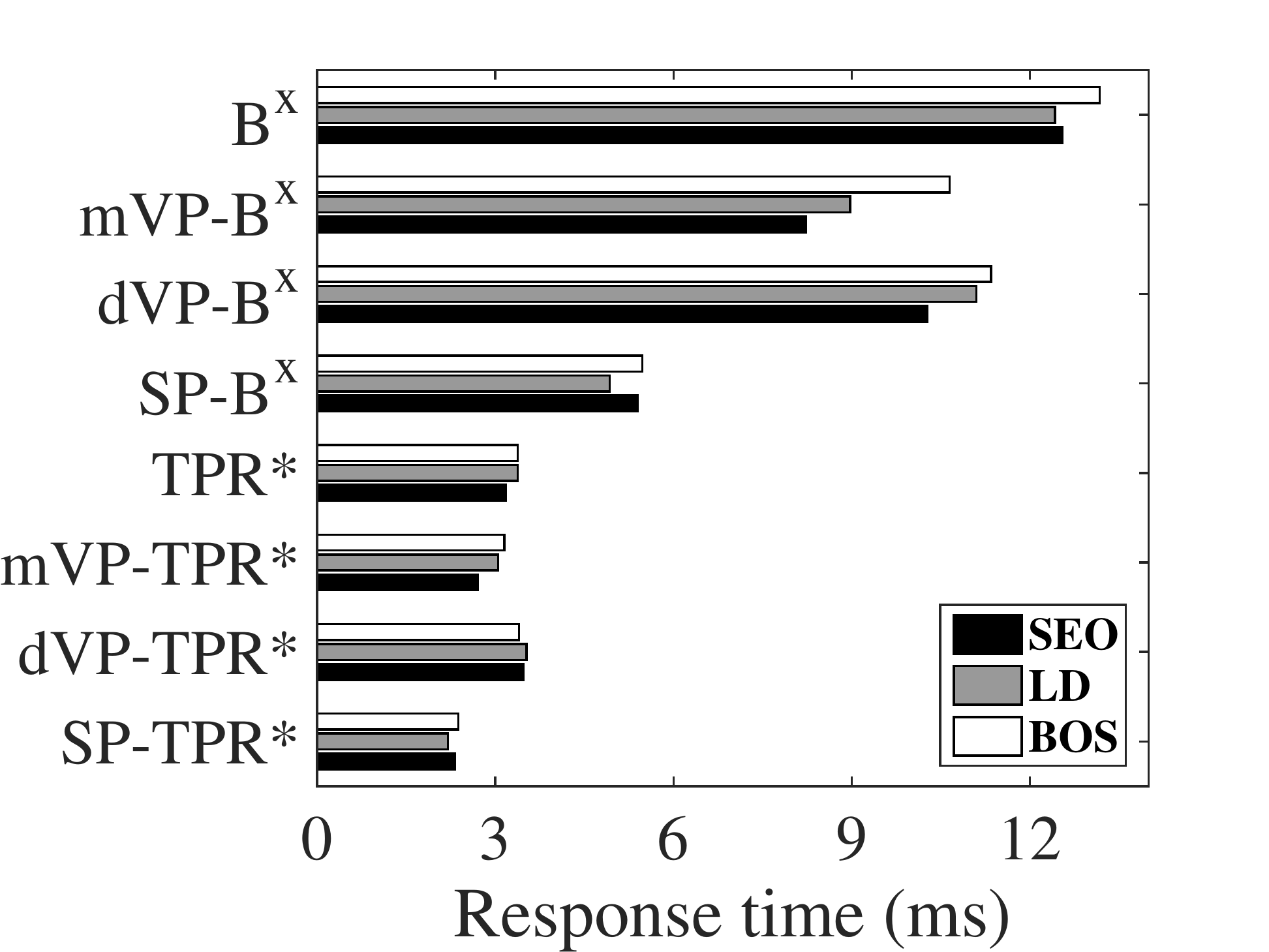}
		\end{minipage}
		\label{da_range}
	}
	\subfigure[$k$NN query]{
		\centering
		\begin{minipage}[b]{0.22\textwidth}
			\includegraphics[width=1.1\textwidth,height=1in]{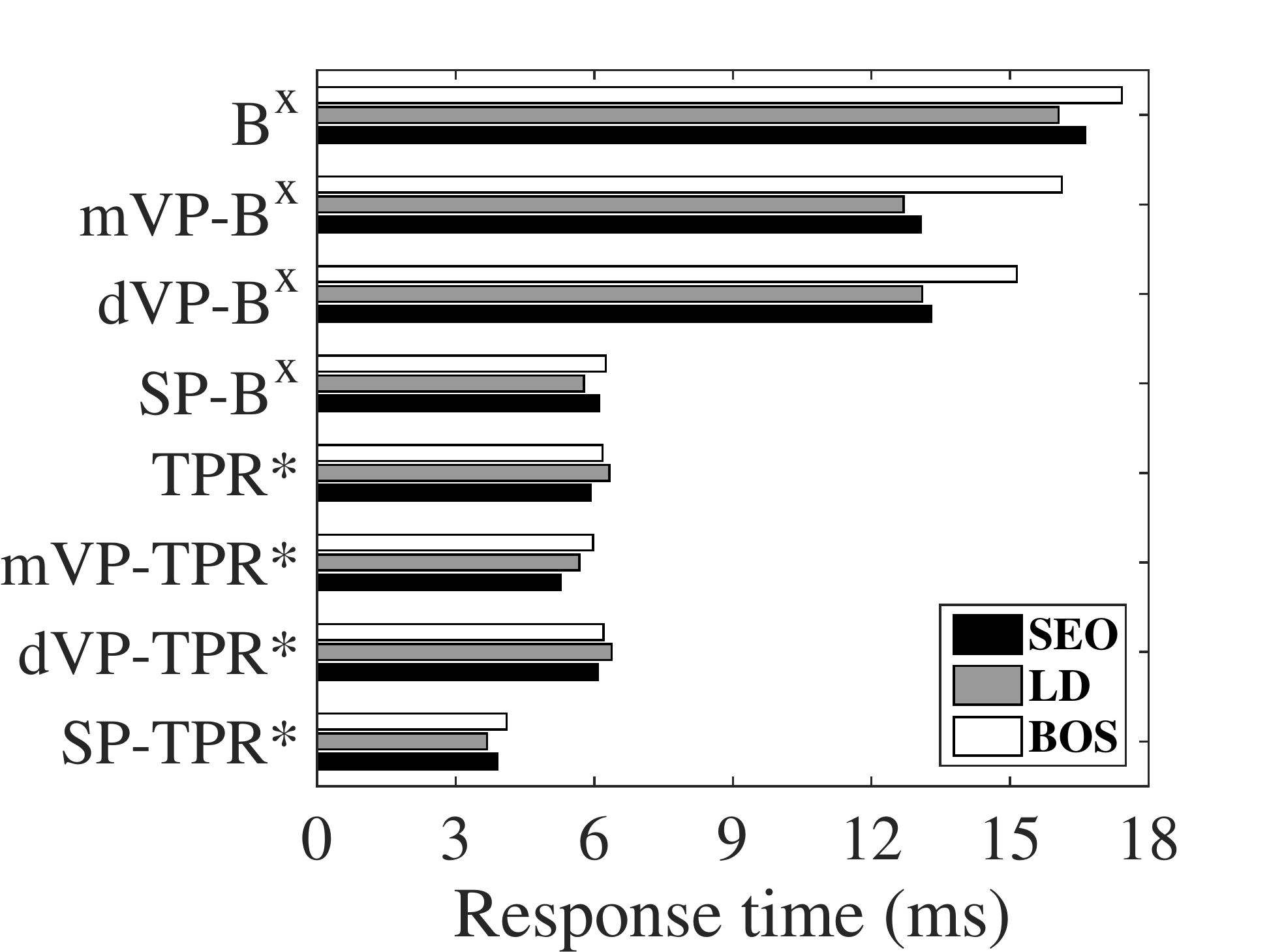}
		\end{minipage}
		\label{da_knn}
	}
	\caption{Vary datasets}
	\label{exp-data}
\end{figure}

Firstly, we show the execution time of Algorithm \ref{algdp} with varied number of objects ($N$) and number of speed values ($q$). Figure \ref{exp-over} shows the results, which are consistent with the complexity analysis for Algorithm \ref{algdp} in Section \ref{sec:part}. We find that the execution time is less then 2 second in all settings. Thus the overhead for partition update is reasonably small. We set $q$ equal 50 in the remaining experiments.

Next we compare the performance difference between disk indexes and main memory indexes. Figure \ref{cs_th} through \ref{cs_knn} show results on throughput, range query response time and $k$NN query response time, respectively. We can see that SP outperforms other methods for both disk and main memory indexes with both B$^x$-trees and TPR$^\star$-trees. Moreover, we found that main memory indexes enjoy much better performance than disk indexes on both throughput and query response time. 
In the remaining experiments, we report only the results of main memory indexes since we have limited space.

Next we compare the experimental results across three simulated traffic data sets (SEO, LD, and BOS), which are summarized in Figure \ref{exp-data}. We can see that SP enjoys better performance than other velocity-based partitioning methods as well as the non-partitioning counterparts on a variety of data sets (road networks from Asian, European, and American cities). This is because, as shown in Figure \ref{network}, road networks for large space domain (10,000$\times$10,000 $m^2$) usually implies no explicit velocity seeds or DVAs which are used in VMBR-based partitioning and DVA-based partitioning techniques, respectively. Moreover, Boston road network has more high speed roads than other city road networks thus nodes in the corresponding indexes expand faster, which makes the BOS data set has higher query costs than other data sets.

\begin{figure}[!h]
	\centering
	\begin{minipage}[b]{0.9\textwidth}
		\includegraphics[height=0.25in]{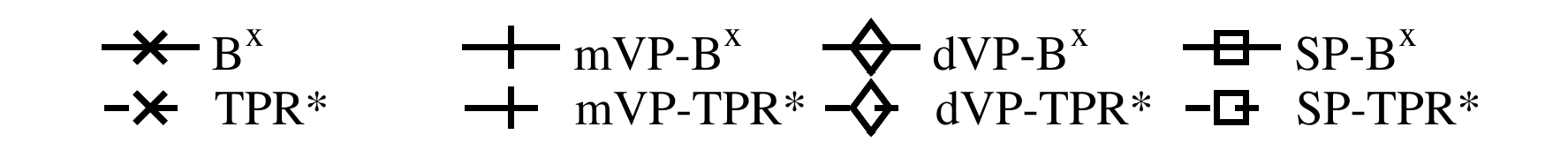}
		\centering
	\end{minipage}
	\subfigure[Throughput]{
		\centering
		\begin{minipage}[b]{0.22\textwidth}
			\includegraphics[width=1.1\textwidth,height=1in]{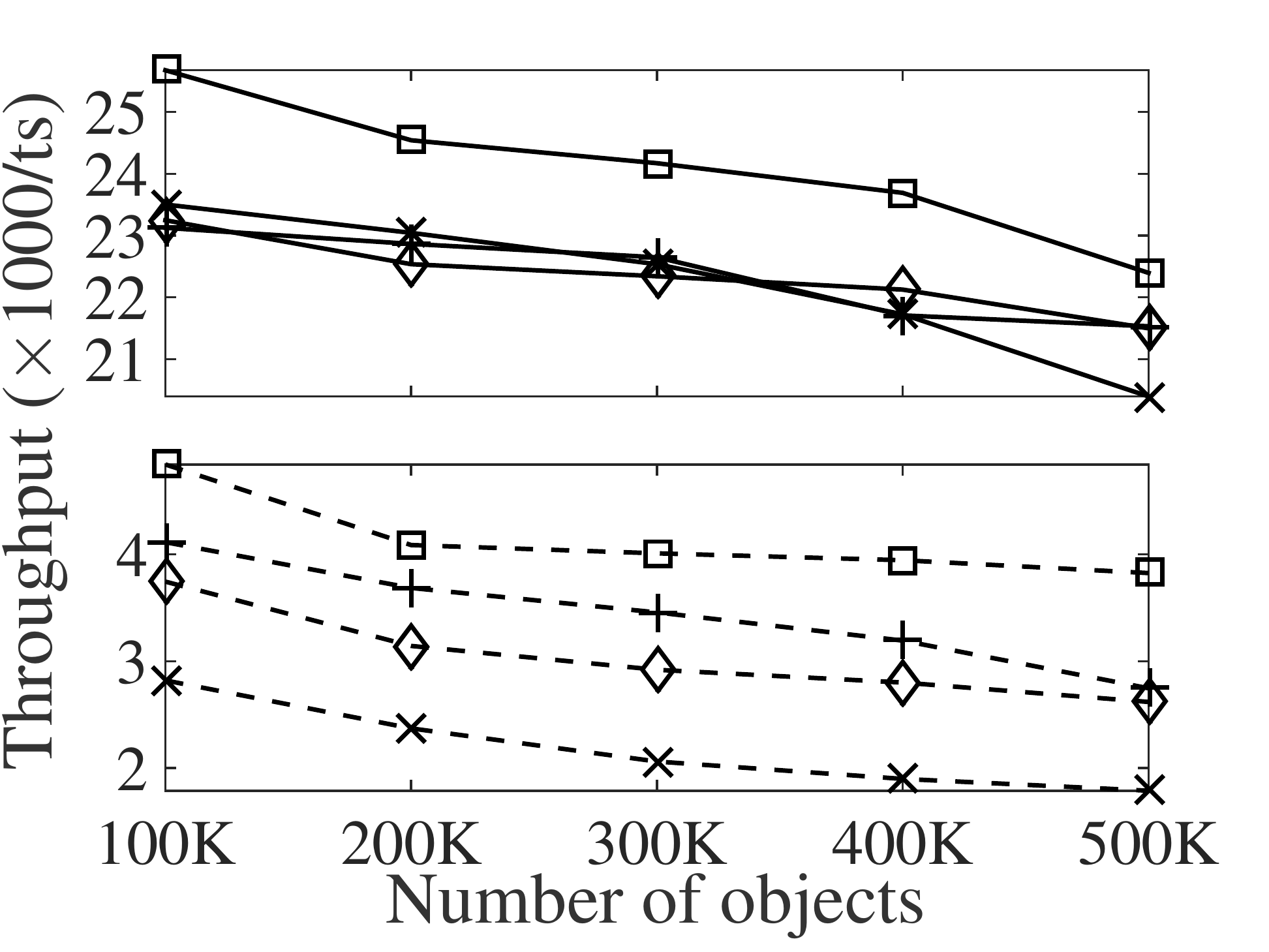}
		\end{minipage}
		\label{no_th}
	}
	\subfigure[Range query]{
		\centering
		\begin{minipage}[b]{0.22\textwidth}
			\includegraphics[width=1.1\textwidth,height=1in]{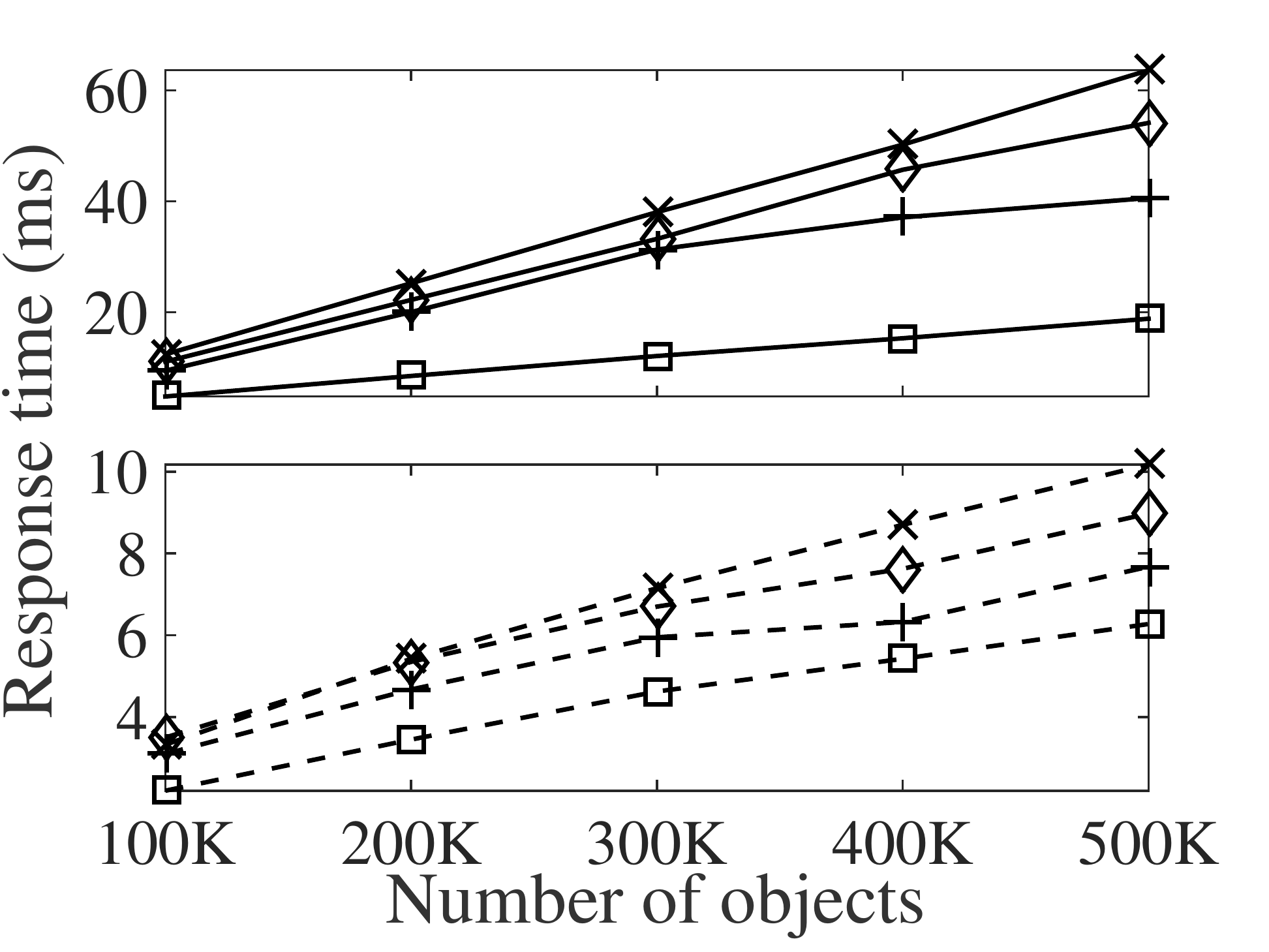}
		\end{minipage}
		\label{no_range}
	}
	\subfigure[$k$NN query]{
		\centering
		\begin{minipage}[b]{0.22\textwidth}
			\includegraphics[width=1.1\textwidth,height=1in]{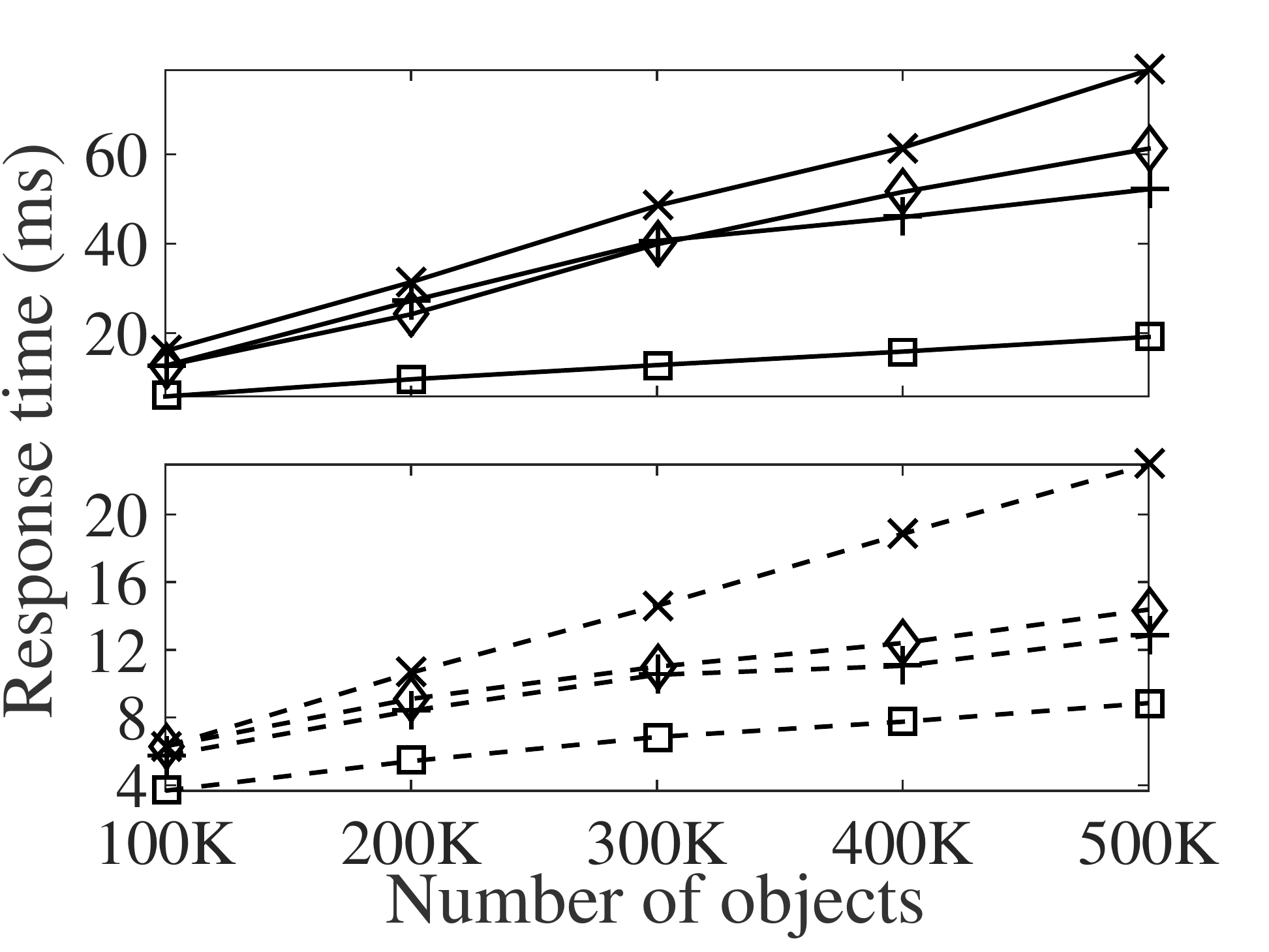}
		\end{minipage}
		\label{no_knn}
	}
	\caption{Vary number of objects}
	\label{exp-no}
\end{figure}

In the next experiment, we vary the number of moving objects from 100K to 500K. Figure \ref{exp-no} shows the results about throughput, range query and $k$NN query. We can see that when the number of objects increases, throughput decreases, query response time increases for both range queries and $k$NN queries. Moreover, B$^x$-trees enjoy higher throughput due to the simple update process of B$^+$-tree but lower query utility due to the ``false hits" caused by the space-filling curves \cite{DBLP:conf/vldb/JensenLO04} \cite{DBLP:journals/vldb/YiuTM08}. On the contrary, TPR$^\star$-trees have more complicated update operations which makes query more efficient at a sacrifice of throughput. Finally, SP indexes consistently outperform other indexes in all settings.

\begin{figure}[!tp]
	\centering
	\subfigure[Throughput]{
		\centering
		\begin{minipage}[b]{0.22\textwidth}
			\includegraphics[width=1.1\textwidth,height=1in]{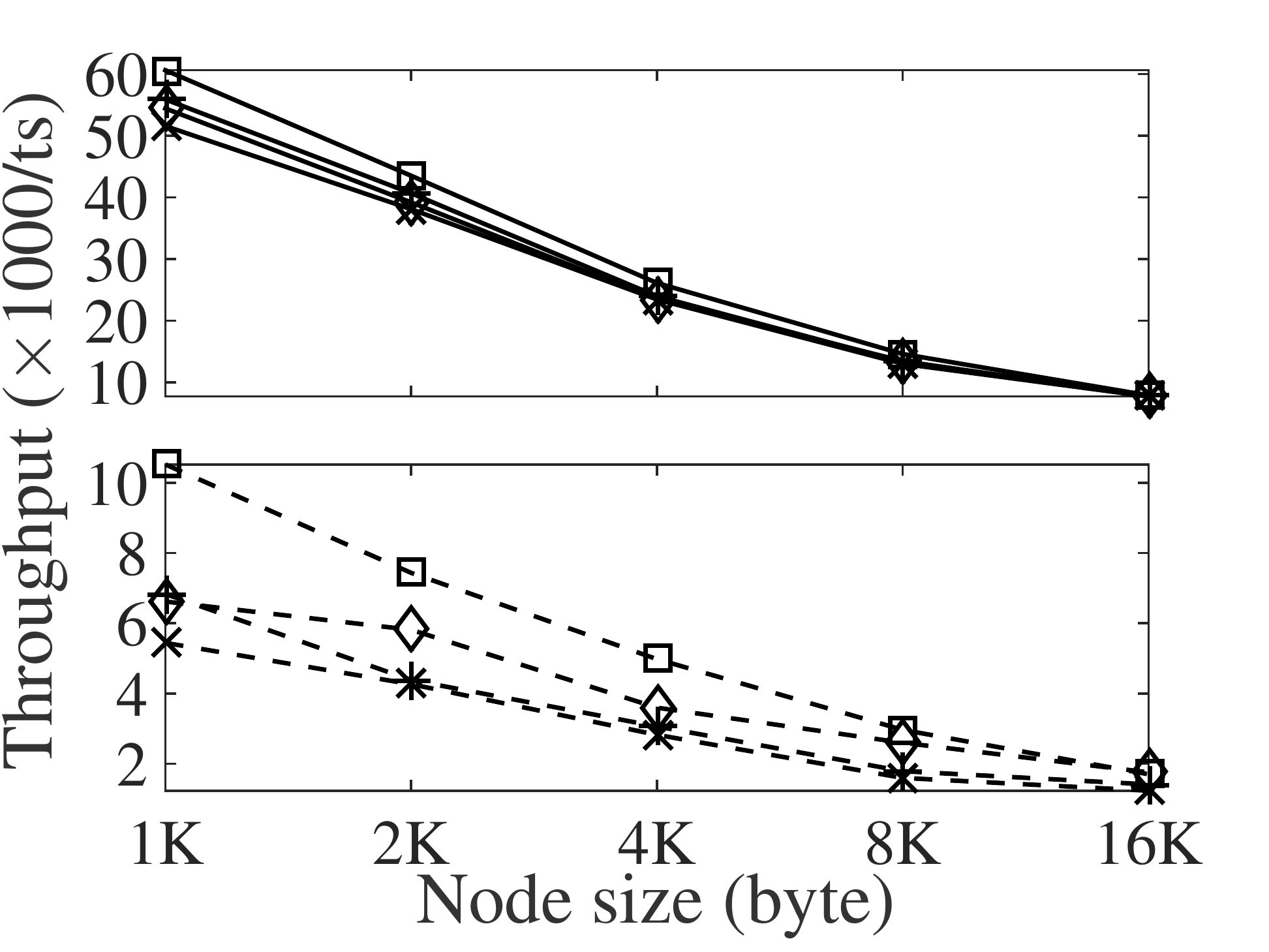}
		\end{minipage}
		\label{nd_th}
	}
	\subfigure[Range query]{
		\centering
		\begin{minipage}[b]{0.22\textwidth}
			\includegraphics[width=1.1\textwidth,height=1in]{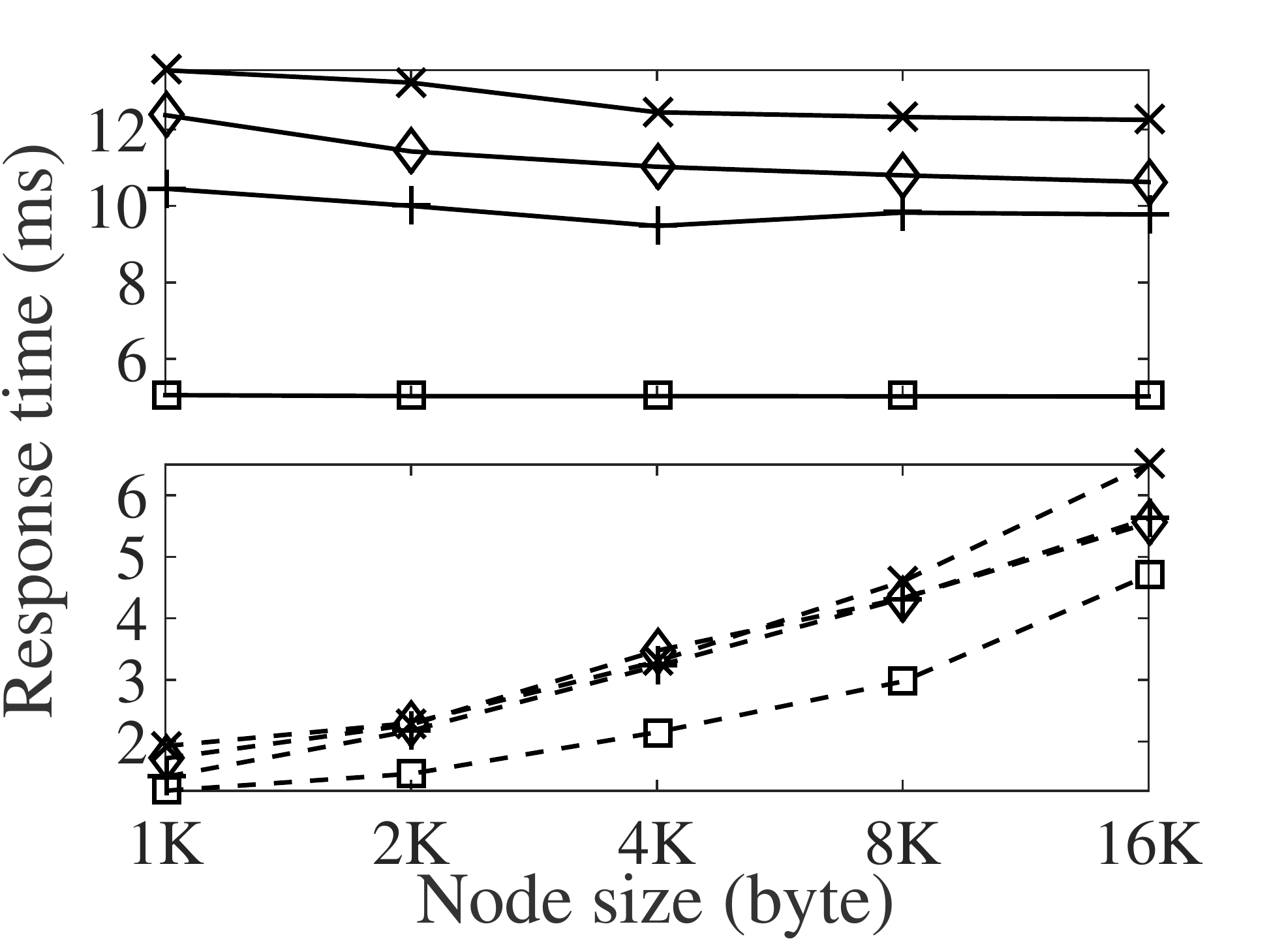}
		\end{minipage}
		\label{nd_range}
	}
	\subfigure[$k$NN query]{
		\centering
		\begin{minipage}[b]{0.22\textwidth}
			\includegraphics[width=1.1\textwidth,height=1in]{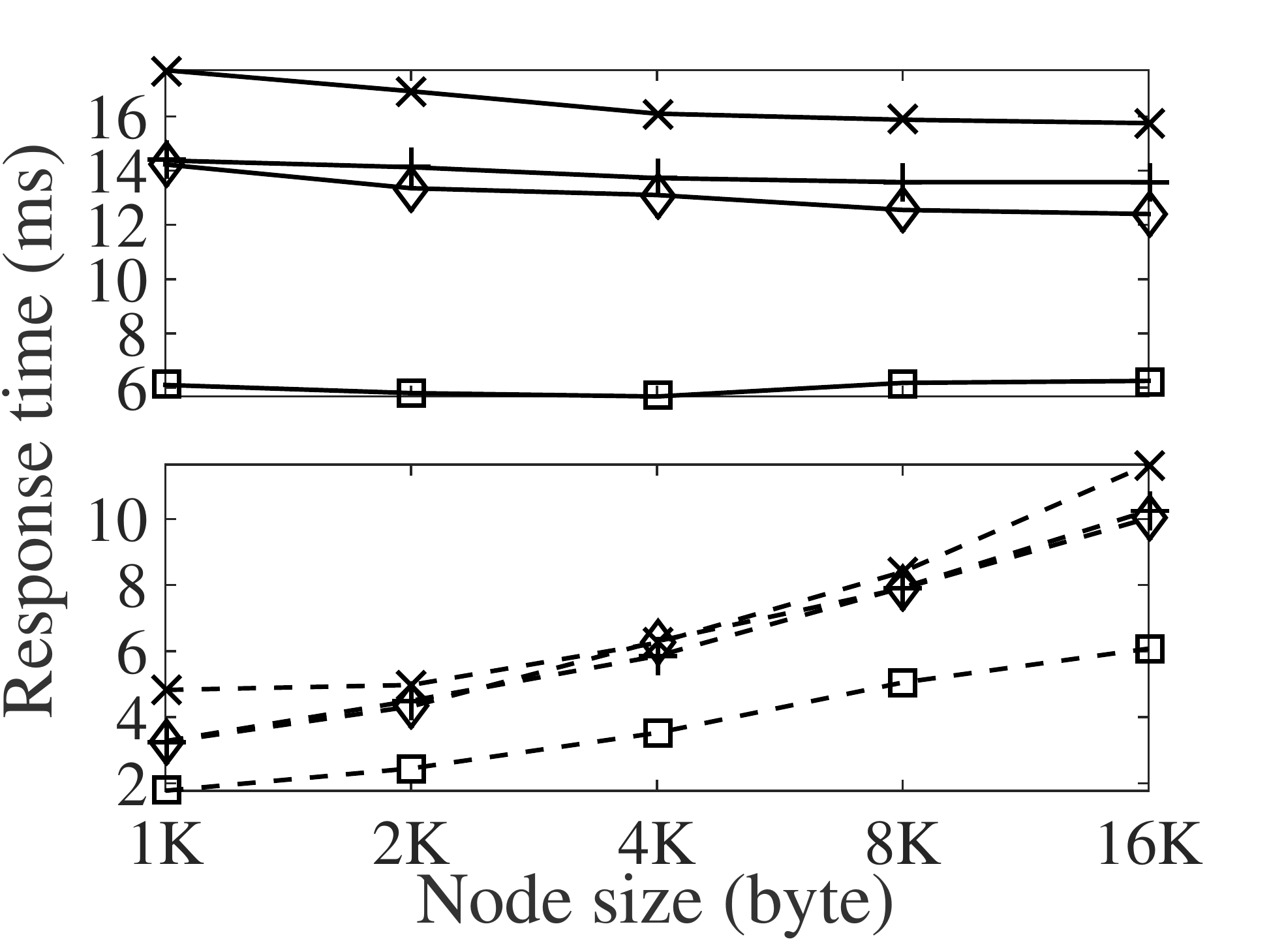}
		\end{minipage}
		\label{nd_knn}
	}
	\caption{Vary node size}
	\label{exp-nd}
\end{figure}

\begin{figure}[!tp]
	\centering
	\subfigure[Range query]{
		\centering
		\begin{minipage}[b]{0.22\textwidth}
			\includegraphics[width=1.1\textwidth,height=1in]{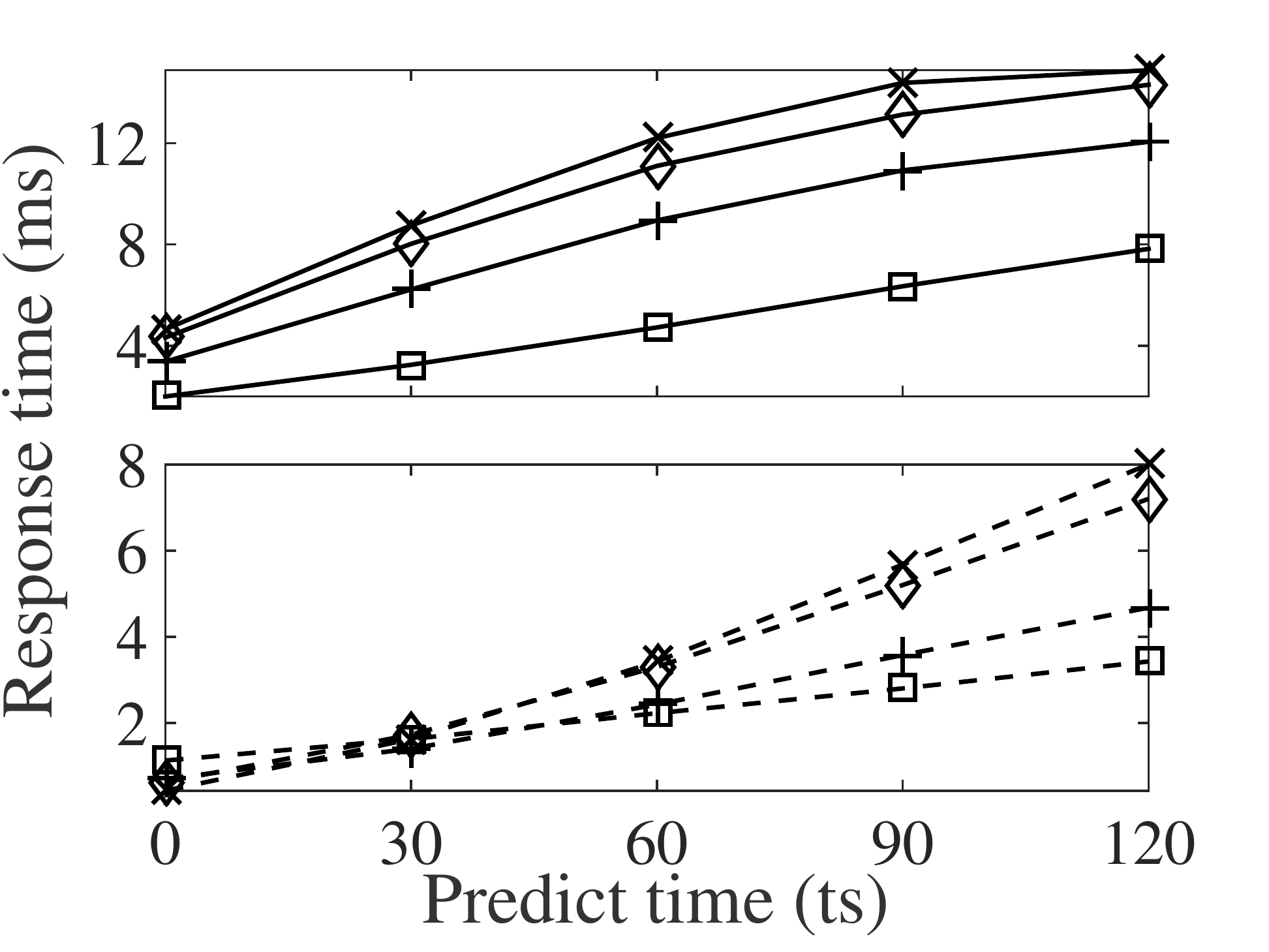}
		\end{minipage}
		\label{pt_range}
	}
	\subfigure[Range query]{
		\centering
		\begin{minipage}[b]{0.22\textwidth}
			\includegraphics[width=1.1\textwidth,height=1in]{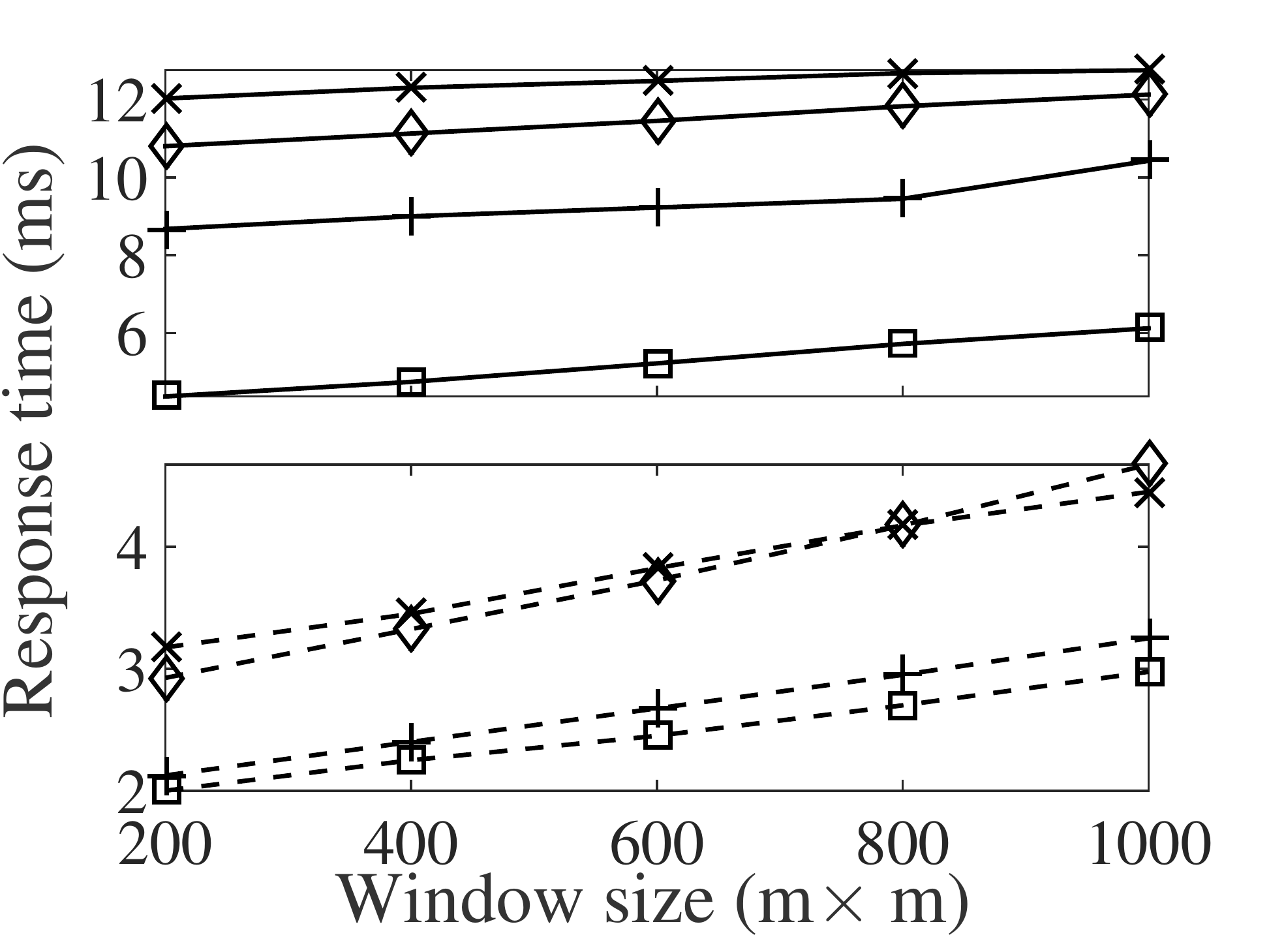}
		\end{minipage}
		\label{ws_range}
	}
	\subfigure[$k$NN query]{
		\centering
		\begin{minipage}[b]{0.22\textwidth}
			\includegraphics[width=1.1\textwidth,height=1in]{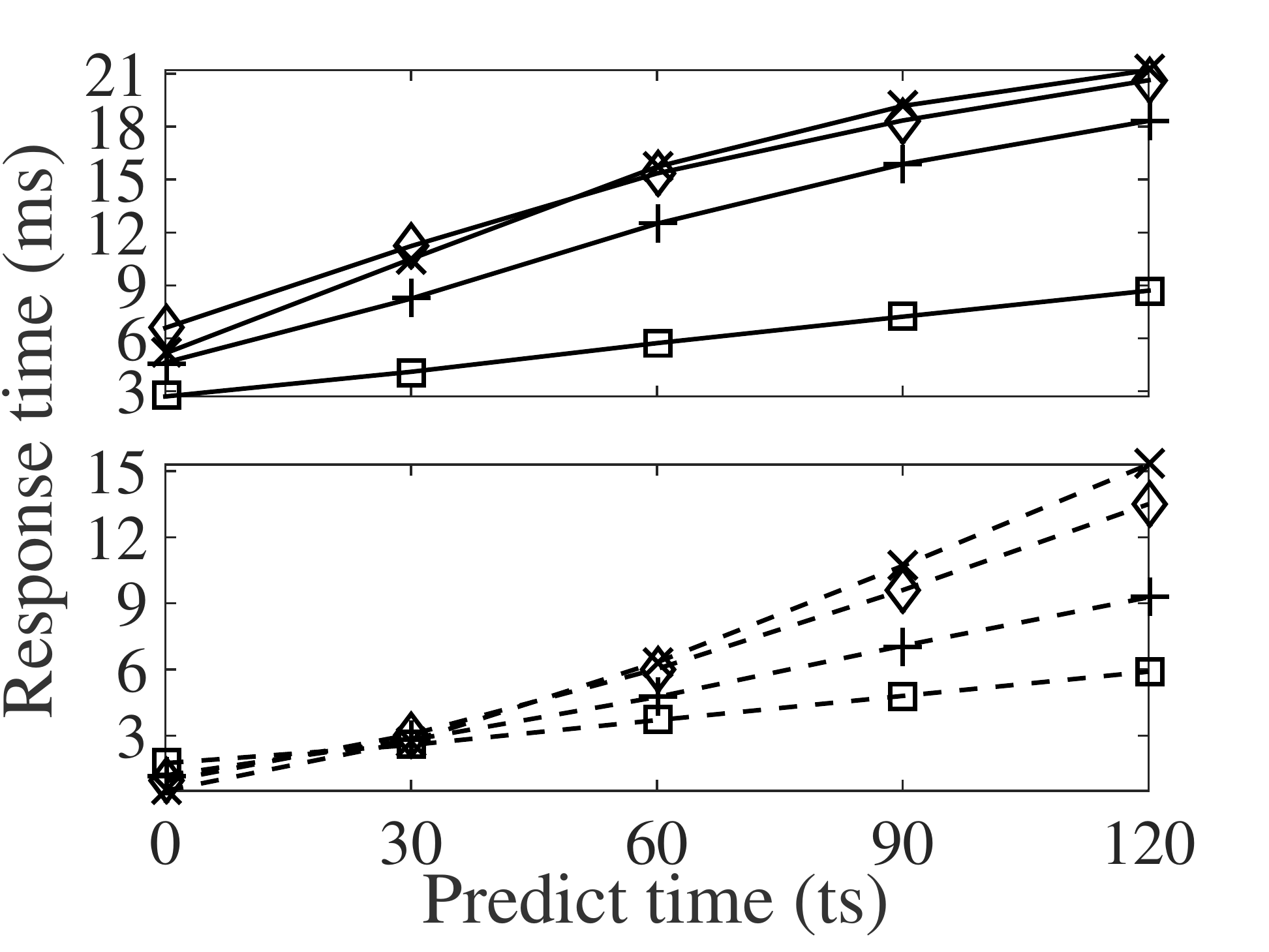}
		\end{minipage}
		\label{pt_knn}
	}
	\subfigure[$k$NN query]{
		\centering
		\begin{minipage}[b]{0.22\textwidth}
			\includegraphics[width=1.1\textwidth,height=1in]{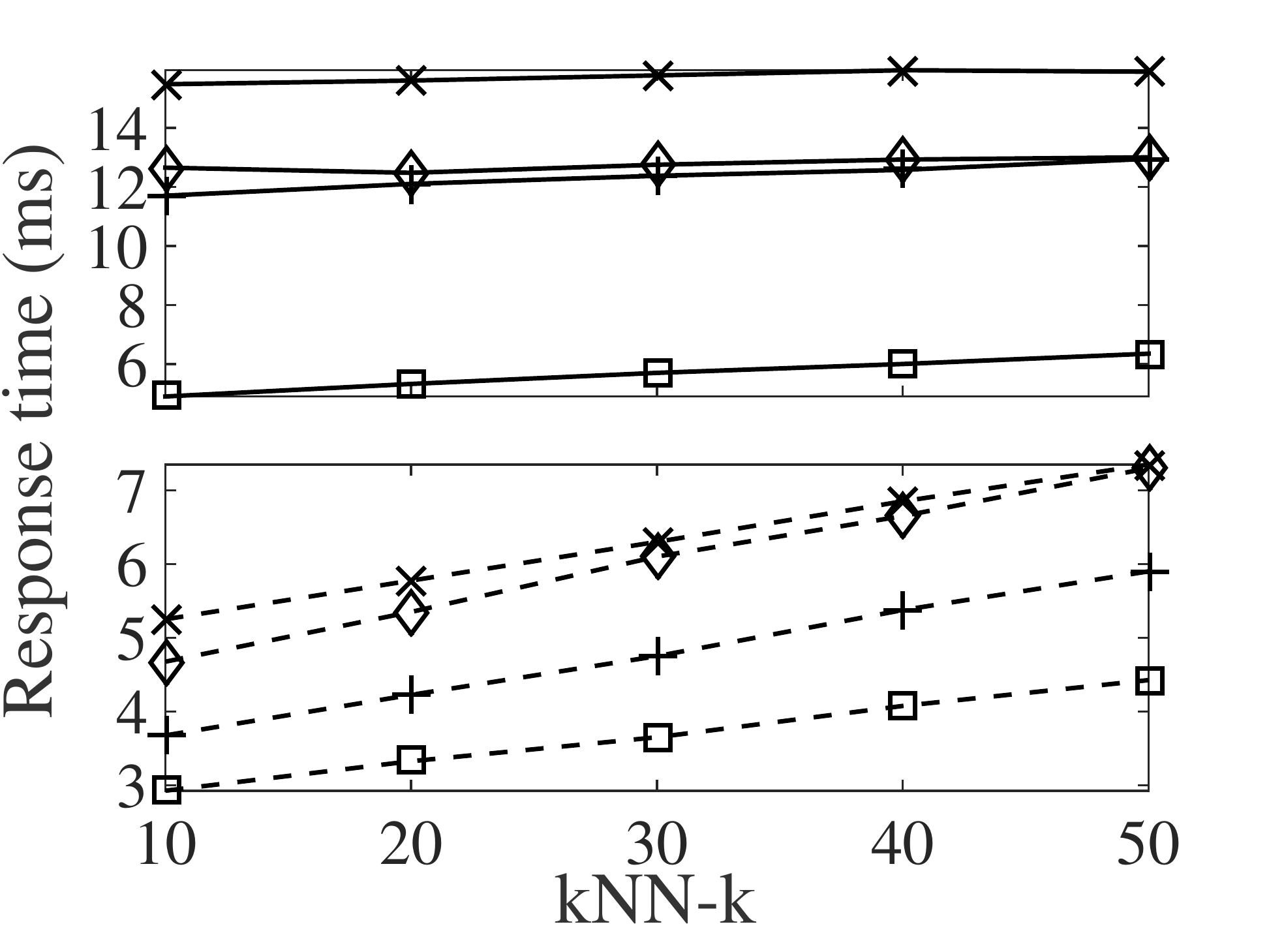}
		\end{minipage}
		\label{knnk_knn}
	}
	\caption{Vary query parameters}
	\label{exp-query}
\end{figure}

\begin{figure}[!tp]
	\centering
	\subfigure[Throughput]{
		\centering
		\begin{minipage}[b]{0.3\textwidth}
			\includegraphics[width=1.1\textwidth,height=1.5in]{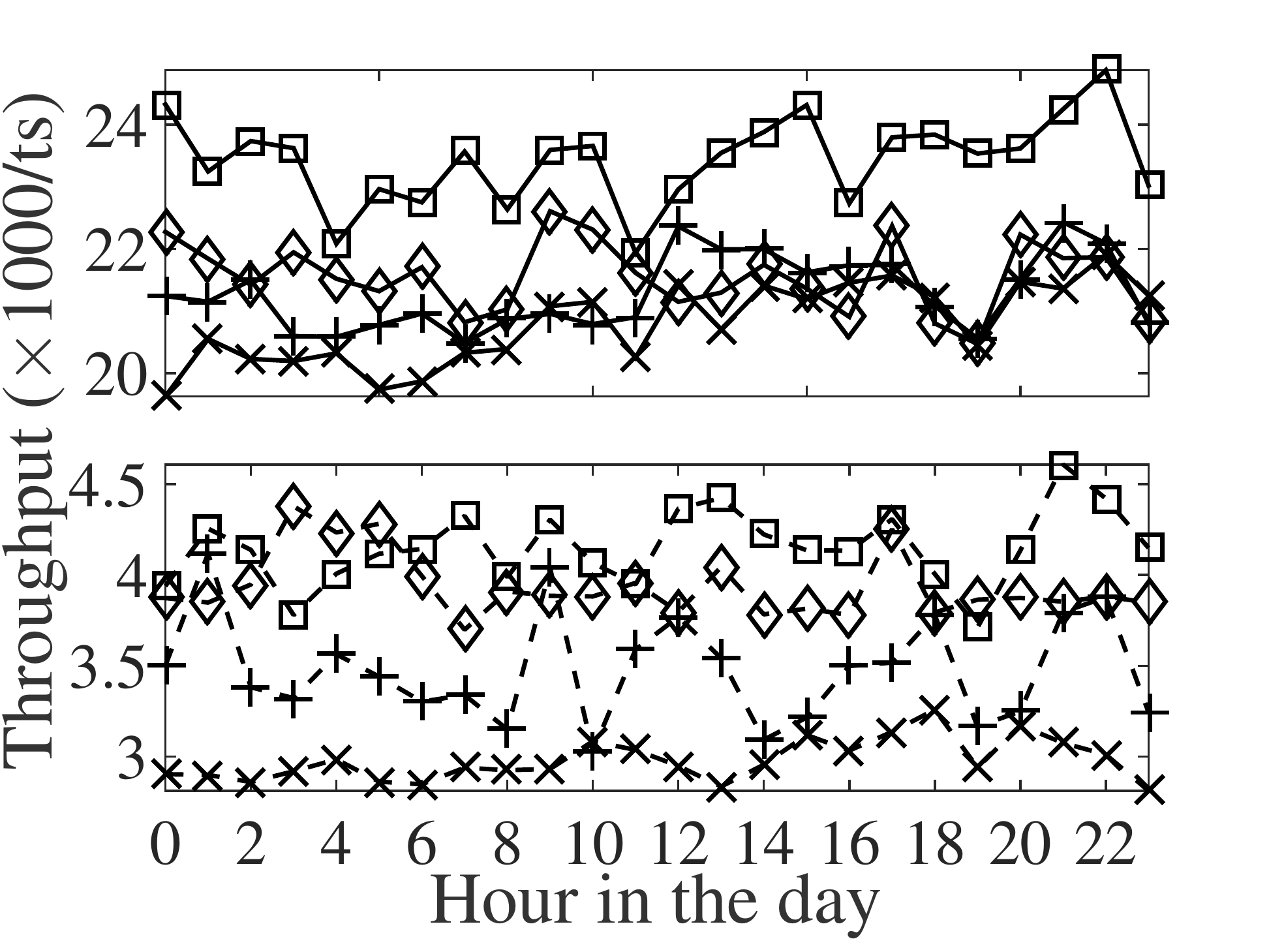}
		\end{minipage}
		\label{gps_th}
	}
	\subfigure[Range query]{
		\centering
		\begin{minipage}[b]{0.3\textwidth}
			\includegraphics[width=1.1\textwidth,height=1.5in]{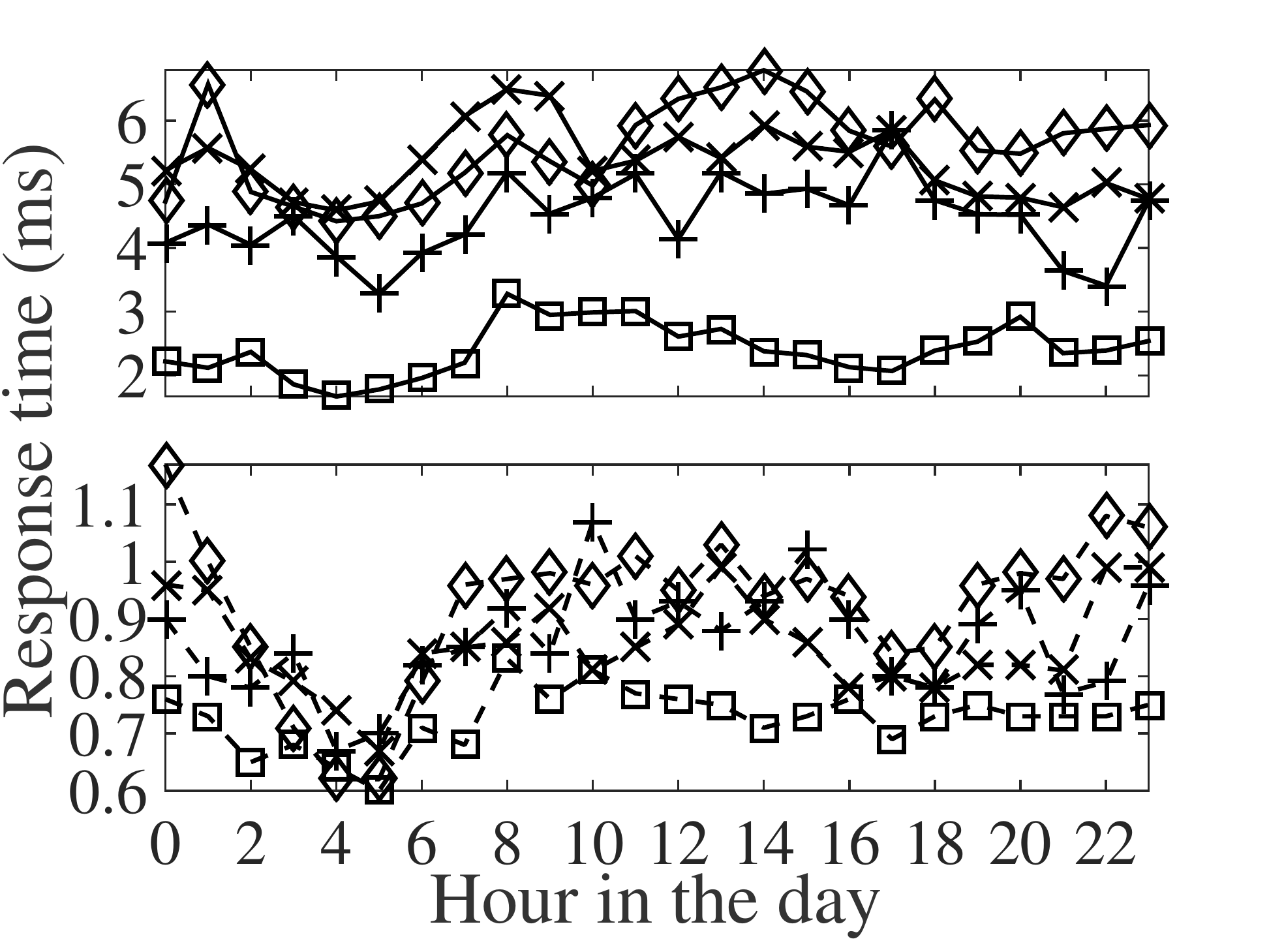}
		\end{minipage}
		\label{gps_range}
	}
	\subfigure[$k$NN query]{
		\centering
		\begin{minipage}[b]{0.3\textwidth}
			\includegraphics[width=1.1\textwidth,height=1.5in]{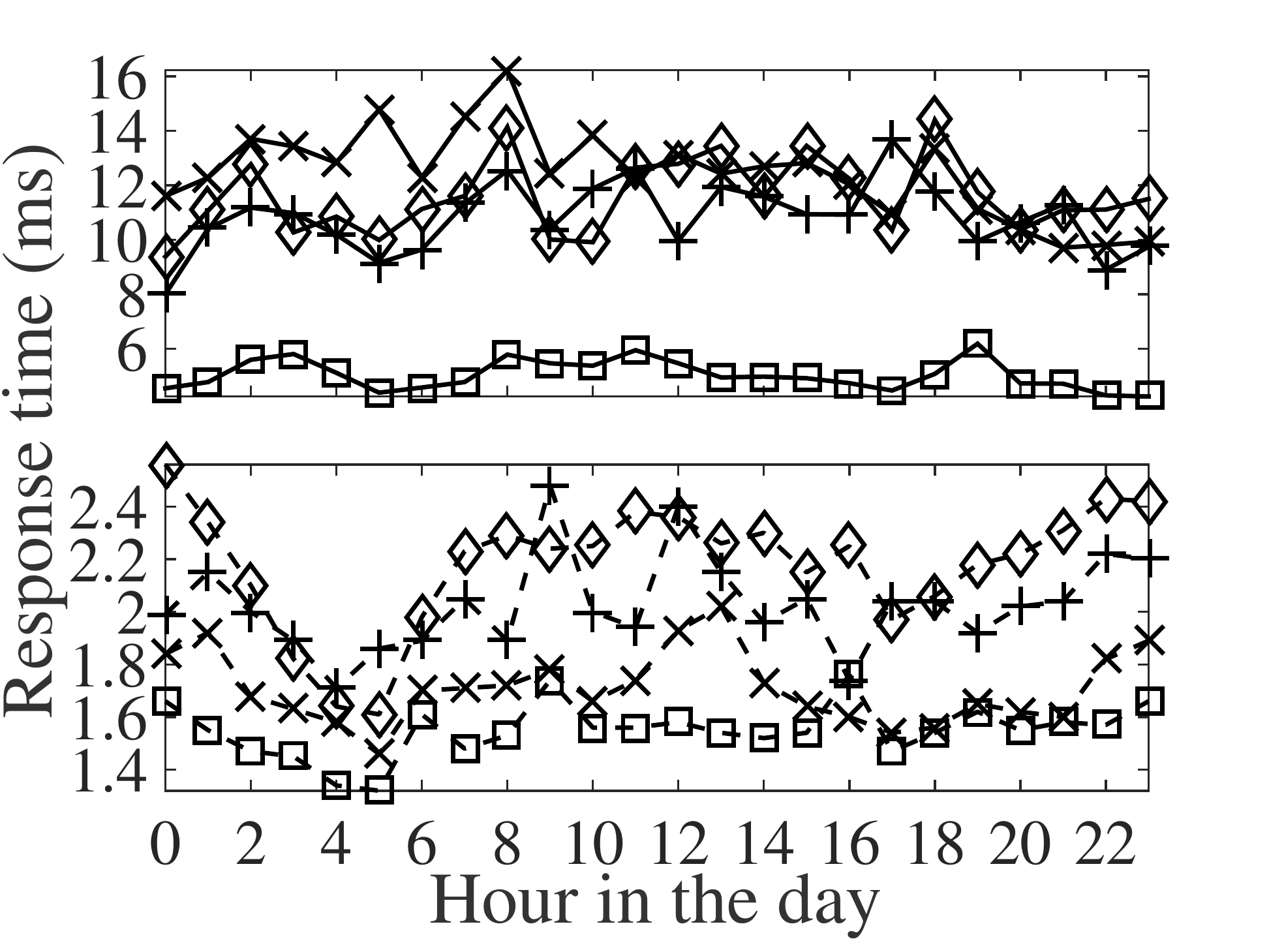}
		\end{minipage}
		\label{gps_knn}
	}
	\caption{Vary hour of the day}
	\label{exp-gps}
\end{figure}

Next we vary the node size from 1KB to 16KB. Figure \ref{nd_th} through \ref{nd_knn} show the experimental results. Generally speaking, performance decreases when node size increases, since index nodes with larger sizes require more maintaining and retrieving efforts. However, query performance of B$^x$-trees is not significantly affected by node size. This is because nodes of B$^x$-trees store the values computed from space-filling curves, which makes the spatial areas of B$^x$-tree nodes insensitive to their storage sizes. Note that the experimental results are different from both those for disk indexes, where disk I/O latency dominates the performance \cite{DBLP:journals/pvldb/ChenJL08}, and those for main memory indexes with secondary index on object IDs, which enables constant time locating the objects for updates \cite{DBLP:conf/gis/SidlauskasSCJS09}. Finally, SP significantly outperforms other methods in this experiment.

Next we study the impact of query parameters including query predict time, range query window size and $k$NN-$k$. The experimental results are summarized in Figure \ref{exp-query}. Figure \ref{pt_range} and \ref{ws_range} show the results about range queries while Figure \ref{pt_knn} and \ref{knnk_knn}  show those about $k$NN queries. We can conclude from the figures that, generally speaking, TPR$^\star$-trees perform better than B$^x$-trees and SP outperforms other methods. 

Finally, we present the results on the real world data set SZ and show the impact of changing speed and location distributions caused by the time of the day. The SZ data set contains information of the taxis in a day long period, thus the location and speed distributions might change during the experiment time. Thus we perform partition updates every 1 hour. The experimental results are summarized in Figure \ref{exp-gps}. We can see that query costs are lowest at early morning, since most cities have least volume of traffic during that time period. We also find that query costs raise at noon and night. This is because the taxis drive faster resulting in higher expanding speeds of the index nodes. The variation of throughput during the day is relatively small. Again, SP significantly and consistently outperforms other partitioning methods and their unpartitioned counterparts.

\section{Conclusions and Future Work}
\label{sec:conc}
In this paper, we proposed a novel and generic speed partitioning technique (SP) for indexing moving objects and implemented SP with the state-of-the-art indexing structures including the B$^x$-tree and the TPR$^\star$-tree. We empirically evaluated the performance of SP through extensive experiments on both simulated traffic data and real world GPS tracking data.

There are several future works which can further improve the performance of SP. Firstly, seeking more accurate estimations on search space expansion can always help finetune the optimal partitioning. Secondly, analytic methods such as \textit{kernel density estimation} (KDE), instead of empirical methods, can be used to estimate the speed distribution. Moreover, sophisticated partition update algorithms might further improve performance in highly dynamic scenarios, where the distributions of location and velocity change frequently. Finally, we will extend our method to grid-based indexing structures.

\section*{Acknowledgments}
This research is supported by the AFOSR DDDAS program under grant FA9550-12-1-0240.
\bibliographystyle{abbrv}
\bibliography{reference}
\end{document}